\documentclass[twocolumn]{aastex701}

\usepackage{graphicx}

\usepackage{amsmath}
\hypersetup{urlcolor=blue}

\newcommand\be{\begin{equation}}
\newcommand\ee{\end{equation}}

\begin{document}

\title{Correlations with Magnetic Activity in the Solar Near-Surface Shear Layer. I. Rotation}

\author[orcid=0000-0003-0172-3713, sname=Rabello Soares, gname=M. Cristina]{M. Cristina Rabello Soares}
\affiliation{W. W. Hansen Experimental Physics Laboratory, Stanford University, Stanford, CA, 94305-4085, USA}
\email[show]{csoares@sun.stanford.edu}
\correspondingauthor{M. Cristina Rabello Soares}

\author[orcid=0000-0002-6163-3472, sname=Basu, gname=Sarbani]{Sarbani Basu}
\affiliation{Department of Astronomy, Yale University, PO Box 208101, New Haven, CT 06520-8101, USA}
\email[]{sarbani.basu@yale.edu}

\author[orcid=0000-0002-0910-459X, sname=Bogart, gname=Richard]{Richard S. Bogart}
\affiliation{W. W. Hansen Experimental Physics Laboratory, Stanford University, Stanford, CA, 94305-4085, USA}
\email[]{rick@sun.stanford.edu}

\begin{abstract}
We used data from the Helioseismic and Magnetic Imager to determine the rotation rate of the near-surface shear layer and its time variation. We applied the ring-diagram analysis technique allowing us to probe the layer between the depths of 1~Mm and 17~Mm. 
We find that the rotation rate increases inwards; it reaches values consistent with those inferred from global helioseismic analyses in the deeper layers, however, there are differences in the rotation rate of the northern and southern hemispheres. We show that the time variation of the rotation rate can be determined even without subtracting the time-averaged rotation rate from each epoch; however, such a subtraction is needed to get the canonical ``torsional oscillation'' signal. We find that even at depths as shallow as 1~Mm, the rotation rate shows the typical torsional oscillation pattern.
The cumulative zonal displacement inferred from the residual flows exhibits a pronounced high-latitude hemispheric asymmetry and varies on solar-cycle timescales; at $75^\circ$ it shows an apparent temporal association with the polar magnetic field.
We find significant correlations between the cumulative displacement and magnetic activity at a subset of latitudes, with multi-year lags: the displacement leads activity by \(\sim 5\) years near \(15^\circ\), whereas at higher latitudes activity leads by \(\sim 4\) years.
At mid to high latitudes, the inferred lags show a hemispheric dependence, with activity tending to lead in the north and lag in the south, suggesting possible hemispheric differences in the timing of cycle evolution and motivating longer time series to test cycle-to-cycle variation.
\end{abstract}


\section{Introduction} \label{sec:intro}

The rotation rate of the Sun increases inward from the surface to a radius of about $0.95\,R_\odot$. This layer is usually referred to as the near-surface shear layer, henceforth NSSL, of the Sun. The first evidence for the NSSL was that emerging active regions rotate faster than the surrounding photosphere \citep{foukal1972}. Later, helioseismic analyses  confirmed the presence of the NSSL \citep{rhodes1990}, and it has since been a subject of many analyses \citep[e.g.][etc.]{ barekat2014, barekat2016, Howeetal2018, BasuAntia2019, antia2022, komm2022, rabello2024}. 

There is considerable ongoing research on the role of the NSSL in the operation of the solar dynamo. It has been argued that this layer has a role in a distributed solar dynamo \citep{brandenburg2005, pipin2011}, and also in the equatorward migration of solar activity belt \citep{brandenburg2005}. Thus, there has been increasing interest in what this layer can do in terms of solar magnetic field generation \citep[][etc.]{dikpati2002, mason2002, kapyla2006, karak2016, paradkar2019, jha2021}. It has also recently been argued that the NSSL is directly responsible for the solar cycle, through a magnetorotational instability\citep{mri}. 
 
While the NSSL has been probed with helioseismic data many times, the rotation rate in the immediate near-surface layers is still uncertain. \citet{Komm2025} looked at the temporal change in the average flow between 2~Mm and 11.6~Mm below the surface using pipeline-produced data by the Global Oscillation Network Group \citep[GONG;][]{gong}, Michelson Doppler Imager \citep[MDI;][]{mdi}, and the Helioseismic and Magnetic Imager \citep[HMI;][]{hmi}; they did not study behaviour closer to the surface. There are shallower results obtained with the time-distance analysis of HMI data \citep{Zhao2014}\footnote{see also http://jsoc.stanford.edu/data/timed/index.html}, but the authors only discuss changes in the rotation rate, and not the rotation rate itself. In this paper, we present the results of a detailed helioseismic study of the NSSL and its change with time. We use data from the HMI instrument alone to avoid instrument-to-instrument systematics and the often {\it ad hoc} corrections that are applied to remove them. We have used the ring-diagram method of helioseismic analysis \citep{hill1988} for our study.  This method allows us to study this layer in detail and, unlike global helioseismic observations, also allows us to examine north-south asymmetries in the layer.  We use custom-created ring diagrams for this work, which allows us better sensitivity towards the surface, as well as two different inversion techniques for the analysis.  This is the first of two papers investigating the properties of the near-surface shear layer. This paper concentrates on the rotation rate and its time variation from a depth of 1~Mm to 17~Mm; the shallowest layer has not been probed in much detail earlier, while the deepest allows us to compare our results with those obtained with global helioseismic analyses.  Paper~II is an examination of the rotational shear, as quantified by the dimensionless radial gradient of the rotation rate, and its correlation with magnetic activity.

The rest of the paper is organized at follows: we discuss details of the data analysis and the inversion techniques in Section~\ref{sec:data}; we present our results in Section~\ref{sec:res}, we discuss the implications of the results and state our conclusions in Section~\ref{sec:conclusion}. 

\section{Data and Analysis}
\label{sec:data}

\subsection{Obtaining the rings}

We use Dopplergrams from the Helioseismic and Magnetic Imager\citep[HMI:][]{hmi}, an instrument on board the Solar Dynamics Observatory (SDO). The data span the time interval from 2010 May 1 (Carrington rotation 2096) through 2026 Jan 24 (Carrington rotation 2307). 
For ring-diagram analyses,  the solar photosphere is tiled with sets of mapped overlapping circular analysis regions in the Dopplergrams centered at fixed Carrington coordinates. For this work we use tiles that have a diameter of  $15$ degrees.  The tiles are tracked for 28.8 hours, about the time it takes to rotate through their diameter at the Carrington rate, using the HMI pipeline module \texttt{mtrack}\citep{Bogart_pipeline}. In this work we only use tiles on the central meridian and the equator. The centers of  $15^\circ$ tiles are separated by $7\fdg 5$.  

Next, the  spatial-temporal power spectra of the resultant data cubes are averaged over a Carrington rotation --- averaging 24 power spectra for rotation. The centers of the $15$-degree tiles scan a range of $\pm 75.0^\circ$ along the central meridian and $\pm 67.5^\circ$ along the equator.

The rotation-averaged spectra were fitted using module \texttt{rdfitc} \citep{Bogart_pipeline} that fits the model of \citet{flow_model} to obtain the shift \(u_x(\theta,t,\ell,n,\nu)\) caused by the zonal (east-west) velocities for each detected mode of degree $\ell$, radial order $n$ and temporal frequency $\nu$. The $u_x$ parameter includes the rotational velocity (modulo the tracking rate) and any other prograde or retrograde zonal flows. 

To study temporal variations and solar-cycle trends, while suppressing short-timescale variations reported by \citet{Bogart2023}, the fitted $u_x$ time series for each $(\ell,n)$ pair at a given latitude $\theta$ were smoothed with a 1-yr (13 Carrington rotations) running mean, advanced by one Carrington rotation. The resulting 1-yr means of the mode parameters were then inverted (see Section~\ref{subsec:inv}). Finally, to obtain overall time averages, we additionally averaged 16 non-overlapping yearly inversion results, $U_x(r,t,\theta)$.

\subsection{Determining the rotation rate}
\label{subsec:inv}

The radial profile of the zonal velocity $U_x(\theta,t,r)$ was obtained by inverting the corresponding one-year averaged fitted  $u_x$. The inversions were carried out using two inversion techniques: Optimally Localized Averages (OLA) \citep[module \texttt{rdvinv};][]{Bogart_pipeline},  and Regularized Least Squares (RLS) using the implementation of \citet{basu_etal1999}.
For each method, we determined the regularization/trade-off parameter for the inversions by balancing noise amplification against localization \citep{rs1999}. 
We use both OLA and RLS inversions as a consistency check; because the methods are complementary \citep[e.g.,][]{sekii1997}, their agreement increases confidence in our inferred results.

The inversion results $U_x$, can be converted to the rotation rate $\Omega$ (in nHz) for each tile and each depth easily using $\Omega=U_x / (r\cos\theta)$ and adding the tracking rate.

\section{Results}
\label{sec:res}

\subsection{Average rotation rate along the central meridian and equator }

\begin{figure}
	\includegraphics[width=\columnwidth]{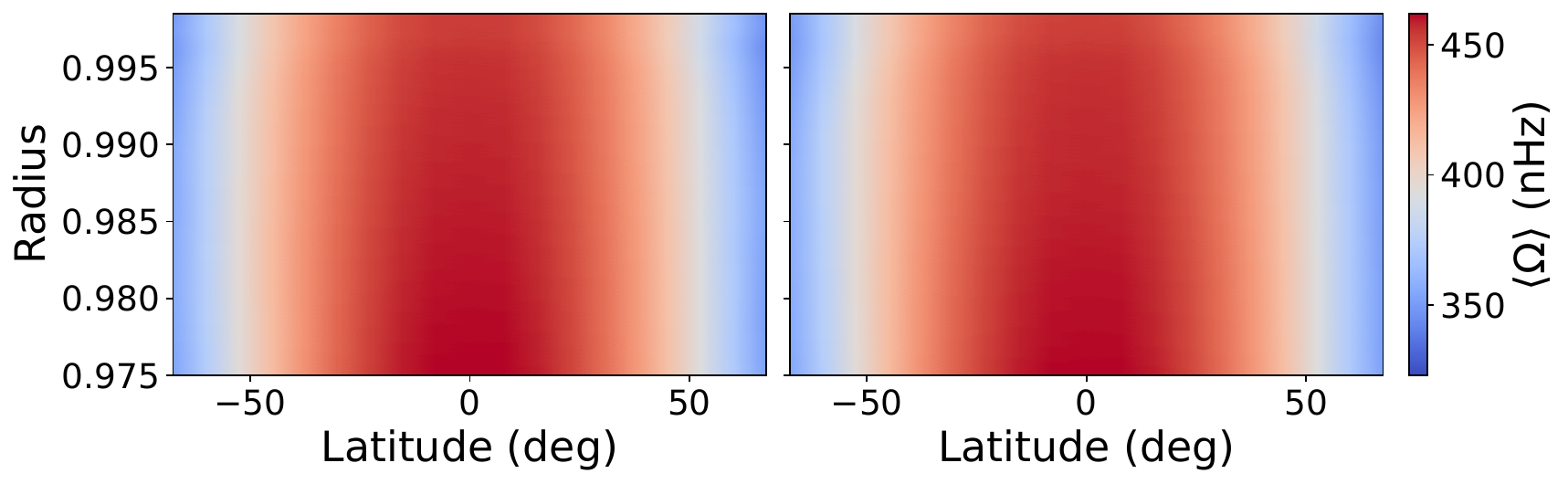}
    \includegraphics[width=\columnwidth]{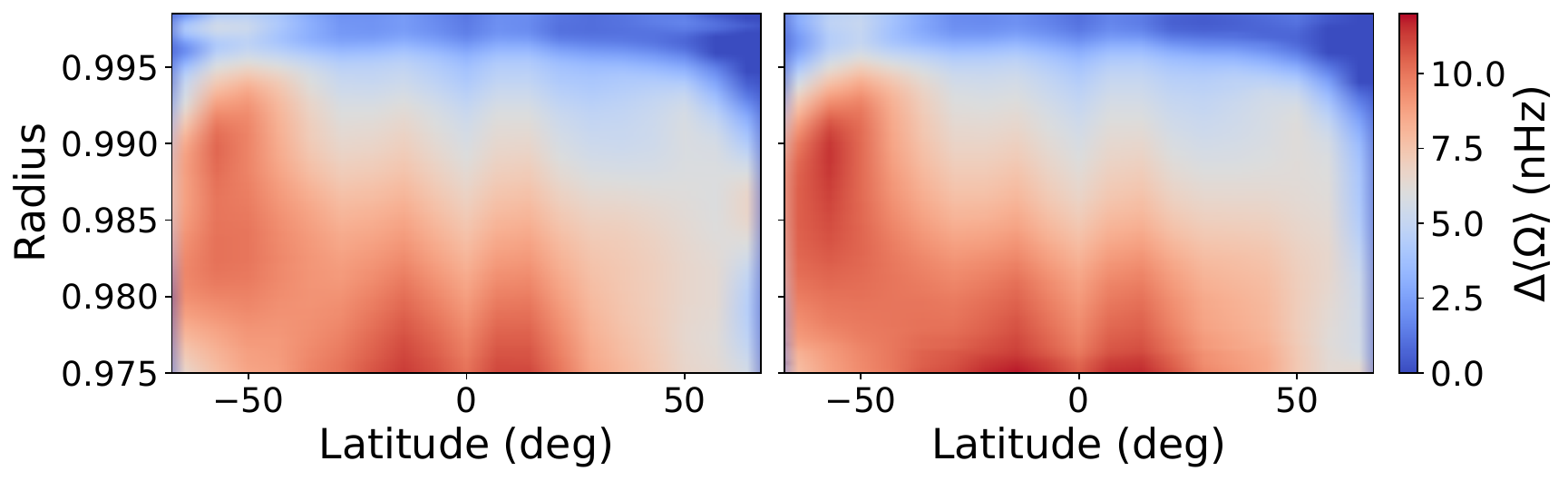}
    \includegraphics[width=\columnwidth]{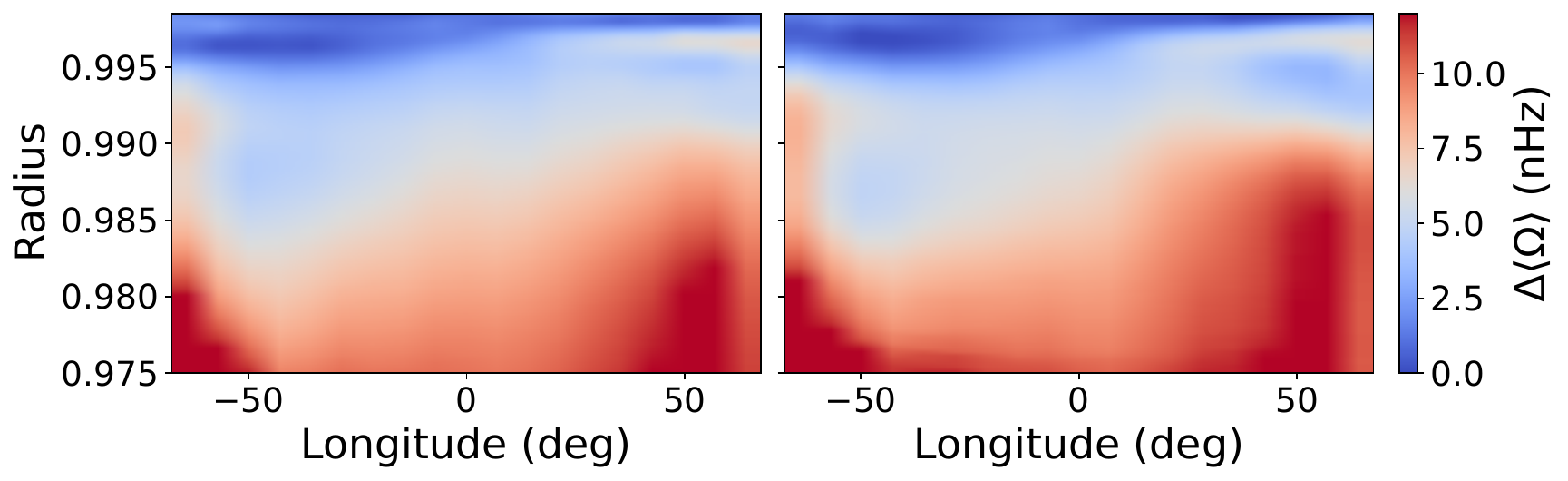}
    \caption{ 
    Top row: The weighted time-averaged rotation rate at the central meridian plotted at different depths as a function of latitude. Middle row: The residuals in the rotation rate 
    after subtraction of the \citet{snodgrass} differential-rotation profile evaluated at each latitude.
    Results are again for the central meridian and plotted as a function of latitude. 
    Bottom row: Residuals of the rotation rate at different depths 
    along the equator, after subtraction of the \citet{snodgrass} profile evaluated at the equator. 
    The residuals are plotted as a function of depth and longitude. Left and right columns show OLA and RLS results, respectively.
    }
    \label{fig:avg_Omega}
\end{figure}

\begin{figure}
	\includegraphics[width=\columnwidth]{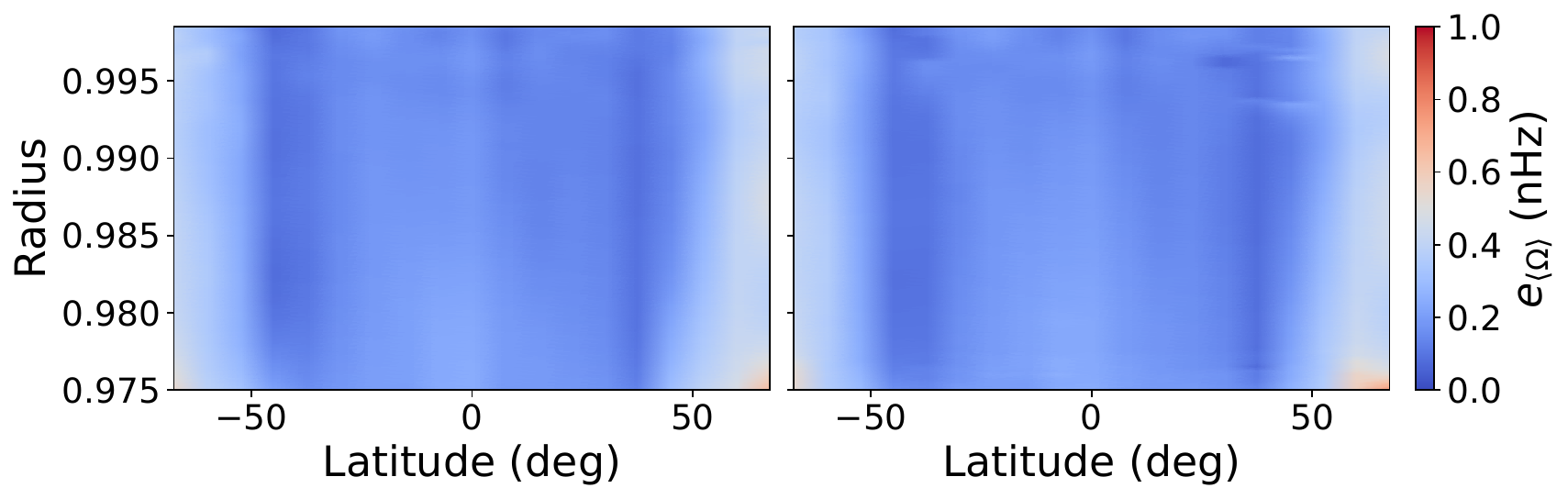}
    \includegraphics[width=\columnwidth]{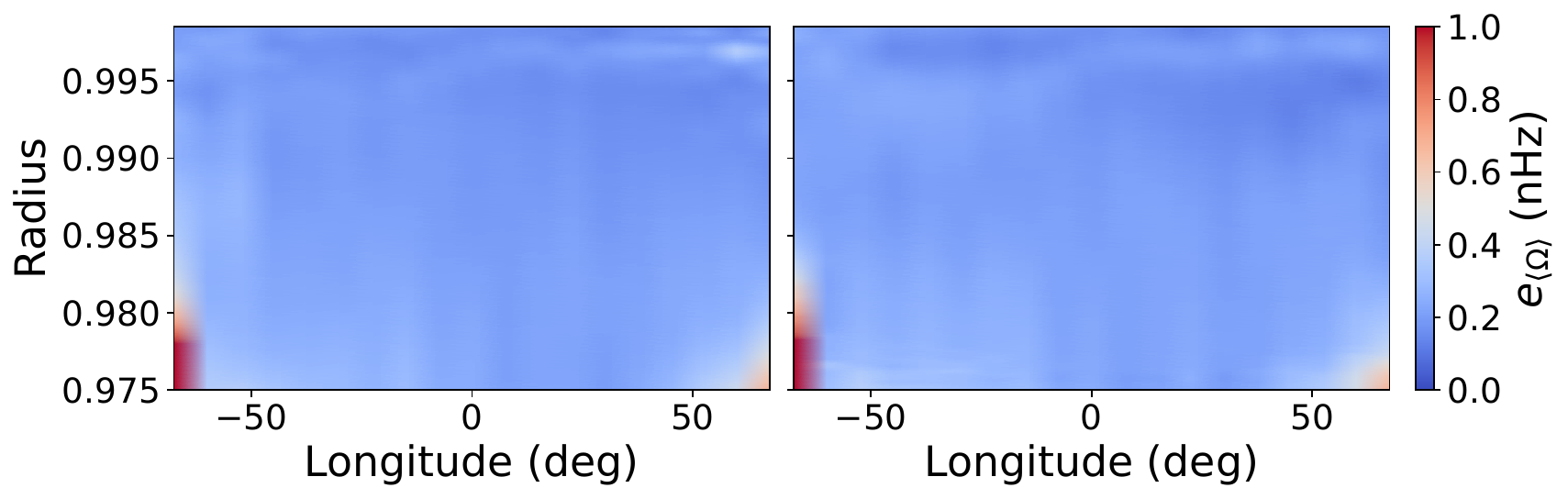}
    \caption{
    Uncertainties in the weighted time-averaged rotation rate. Top row: rotation-rate uncertainties  at the central meridian.  Bottom row: uncertainties at the equator.
    Left and right columns show OLA and RLS results, respectively.}
    \label{fig:avg_eOmega}
\end{figure}

The time-averaged rotation rate in the NSSL is shown in Fig.~\ref{fig:avg_Omega}. The upper panel shows the well-known latitudinal differential rotation at every layer; we show the results only along the central meridian. The uncertainties in the result are shown in the upper panel of Fig.~\ref{fig:avg_eOmega}, where we can see that uncertainties increase at high latitudes --- this is expected because foreshortening of the Doppler signal makes the data noisy. 
The uncertainties are also large very close to the deepest layer that we probe ($0.975\,R_\odot$); this is caused by the fact that we do not have data on high $n$ modes that are sensitive to the deep layers. We also lack sufficiently high-degree ($\ell$) modes to constrain the near-surface layers; therefore, we present inversion results only down to $0.9987\,R_\odot$ (i.e., $\sim 0.90$~Mm beneath the photosphere).

Unfortunately, any latitude-dependent variation of the NSSL is masked by the strong differential rotation. To alleviate this issue, we subtract the surface differential-rotation rate given by  the  \citet{snodgrass} rate, as reported by \citet[Table~1]{Beck2000}:
\begin{equation}
\Omega(\theta)= 451.36 - 54.75\sin^2(\theta) - 80.21\sin^4(\theta)\quad\mbox{nHz},
\label{eq:snodgrass}
\end{equation}
where $\theta$ is the latitude at the midpoint of each tile. The result can be seen in the middle panel of Fig.~\ref{fig:avg_Omega}. From this figure, it is clear that the NSSL has substructure \citep[see also][]{rabello2024}.
The inferred rotation rate is close to the \citet{snodgrass} rate down to $\sim 2$~Mm, and then increases rapidly at greater depths, by as much as $\sim 10$ nHz. 
Also clear is that there is substantial north-south asymmetry, which we shall look into in more detail later in this section. 

There are considerable differences in  the equatorial rotation rate in the east-west direction too, of similar magnitude to the north–south ones, as can be seen in the lowest panel of Fig.~\ref{fig:avg_Omega}; the uncertainties in the results can be seen in the lower panel of Fig.~\ref{fig:avg_eOmega}.

\begin{figure}
	\includegraphics[width=\columnwidth]{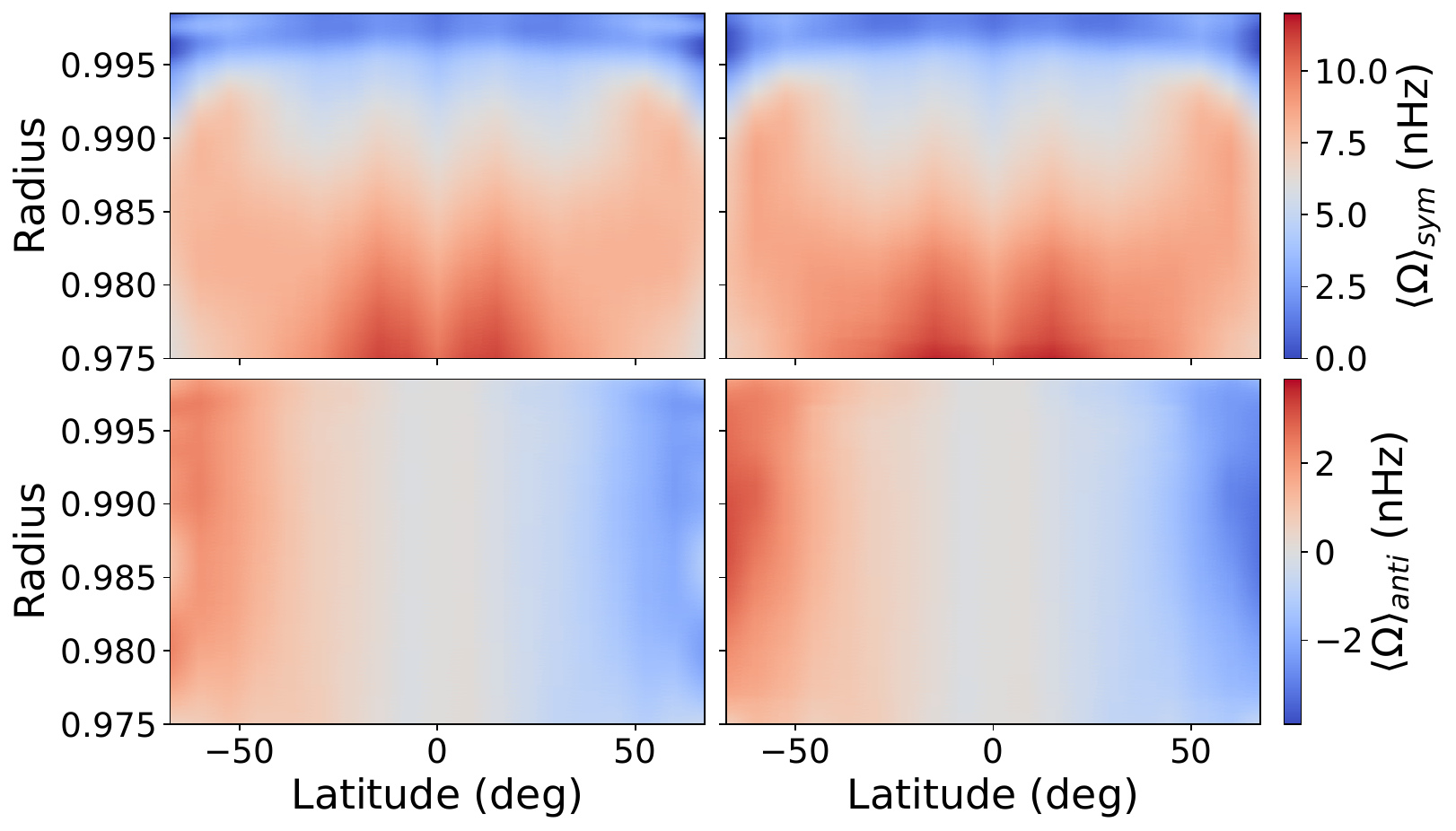}
    \caption{Top: The north-south symmetric component, $[\Omega(N)+\Omega(S)]/2$, of the time-averaged rotation rate plotted as a function of depth for different latitudes. The results are for rings at the central meridian. For the purpose of visualization, we have subtracted 
    the \citet{snodgrass} rate (Eq.~\ref{eq:snodgrass}) at each latitude. 
    Bottom: The north-south antisymmetric component $[\Omega(N)-\Omega(S)]/2$ . The  left and right panels show OLA and RLS results, respectively.
    }
    \label{fig:avg_Omega_nw}
\end{figure}

\subsection{North--south differences}
\label{subsec:ns}

As mentioned earlier, there is a considerable difference between the rotation rate in the north and the south, and this can be seen clearly in Fig.~\ref{fig:avg_Omega_nw}. The southern hemisphere has a slightly higher rotation rate than the northern hemisphere. As we will see later, there is substantial antisymmetry at each epoch; however, we cannot be sure if the antisymmetry in the average rate always holds. We have averaged the rotation rate for Solar Cycle~24 and the rising phase of Cycle~25. It is possible that the picture will change as Cycle~25 progresses and we can average over two full cycles. However, it is well known that solar activity is not north-south symmetric. Data from the Sunspot Index and Long-term Solar Observations \citep[SILSO;][]{SILSO} show that while in the rising phase of Cycle~24, there was an excess of sunspots in the northern hemisphere; the situation reversed at the maximum and reversed again. Cycle~25 started with an excess of sunspots in the southern hemisphere, but that reversed and reversed again at maximum. As we show in Paper~II, the properties of the NSSL correlate with magnetic activity, and hence, the hemispheric asymmetry of the sunspot numbers is relevant here. To determine whether or not the southern hemisphere always rotates faster than the northern hemisphere, we need a much longer dataset. However, it should be noted that the antisymmetry is small. At high latitudes, where the effect is the largest, it is about 1\% of the rotation rate.

\subsection{East--west differences}
\label{subsec:EW}

\begin{figure}
	\includegraphics[width=\columnwidth]{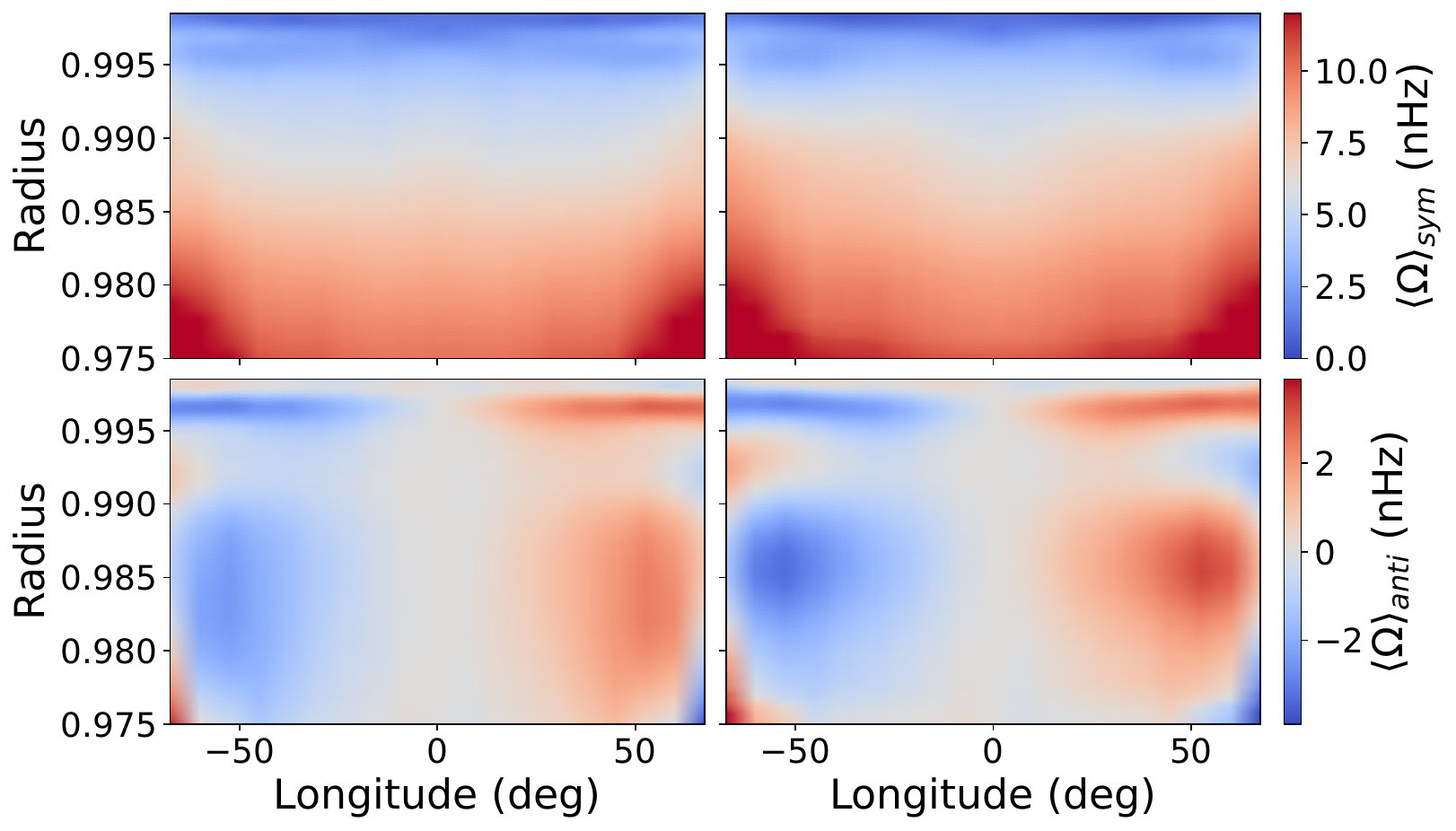}
    \caption{Time-averaged equatorial rotation at the central meridian plotted as a function of depth and longitude. Top: The east-west symmetric about the central meridian, $[\Omega(W)+\Omega(E)]/2$; for visualization purposes, we have subtracted the \citet{snodgrass} profile evaluated at the equator. 
    Bottom: The corresponding antisymmetric component, $[\Omega(W)-\Omega(E)]/2$. The left and right panels are, respectively, results for OLA and RLS inversions.
    }
    \label{fig:avg_Omega_ew}
\end{figure}

\begin{figure}
    \includegraphics[width=\columnwidth]{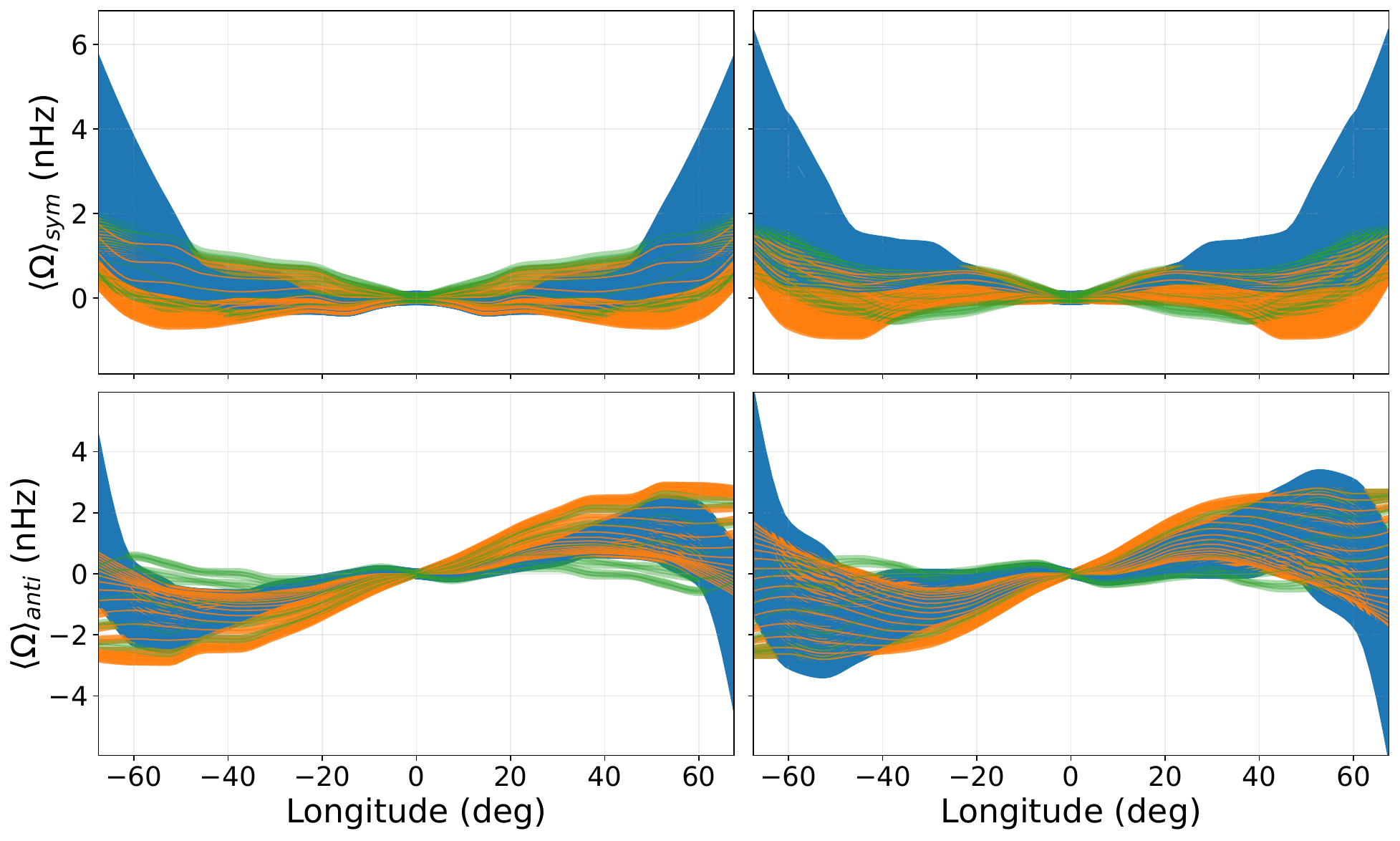}
    \caption{Top: The east-west  symmetric component of the equatorial rotation rate plotted as a function of longitude. For visualization, we have subtracted the center‑meridian value at the same depth. Bottom: The antisymmetric components of the solar rotation rate as functions of latitude at selected depths.
    Color indicates the value range: blue (0.9750–0.9925), orange (0.9925–0.9970), and green (0.9970–0.9988) $R_\odot$. Left and right panels show OLA and RLS inversion results, respectively.
    }
    \label{fig:plot_omega_p2_sym4}
\end{figure}

Our results show an east-west antisymmetry, almost certainly due to spatial variation in the observations and/or analysis, as can be seen in Fig.~\ref{fig:avg_Omega_ew}. Again,  the effect is small, though statistically significant. The effect is seen more clearly in Fig.~\ref{fig:plot_omega_p2_sym4}.

This east-west effect has been seen in local helioseismic results earlier. \citet{zhao2012} showed this effect in time-distance measurements. They called it a ``center-to-limb'' effect.  This effect has defied full explanation thus far. \citet{baldner2012} attributed this effect partially to  the highly asymmetrical nature of the solar granulation (slow, warm upflows that occupy a much larger area than the fast, cold downflows) and that the oscillation modes could react to this as though there is a net radial flow and impart a phase shift on the modes as a function of observing height and thus heliocentric angle. Often {\it ad hoc} corrections \cite[[e.g.,][etc.]{zhao2012, komm2022} are applied to eliminate the ``center-to-limb'' effect. We choose not to apply such a correction since there is as yet no physics-informed basis to correct the data.

\subsection{Time-variation of the NSSL}
\label{subsec:zonal}

\begin{figure}
	\includegraphics[width=0.98\linewidth]{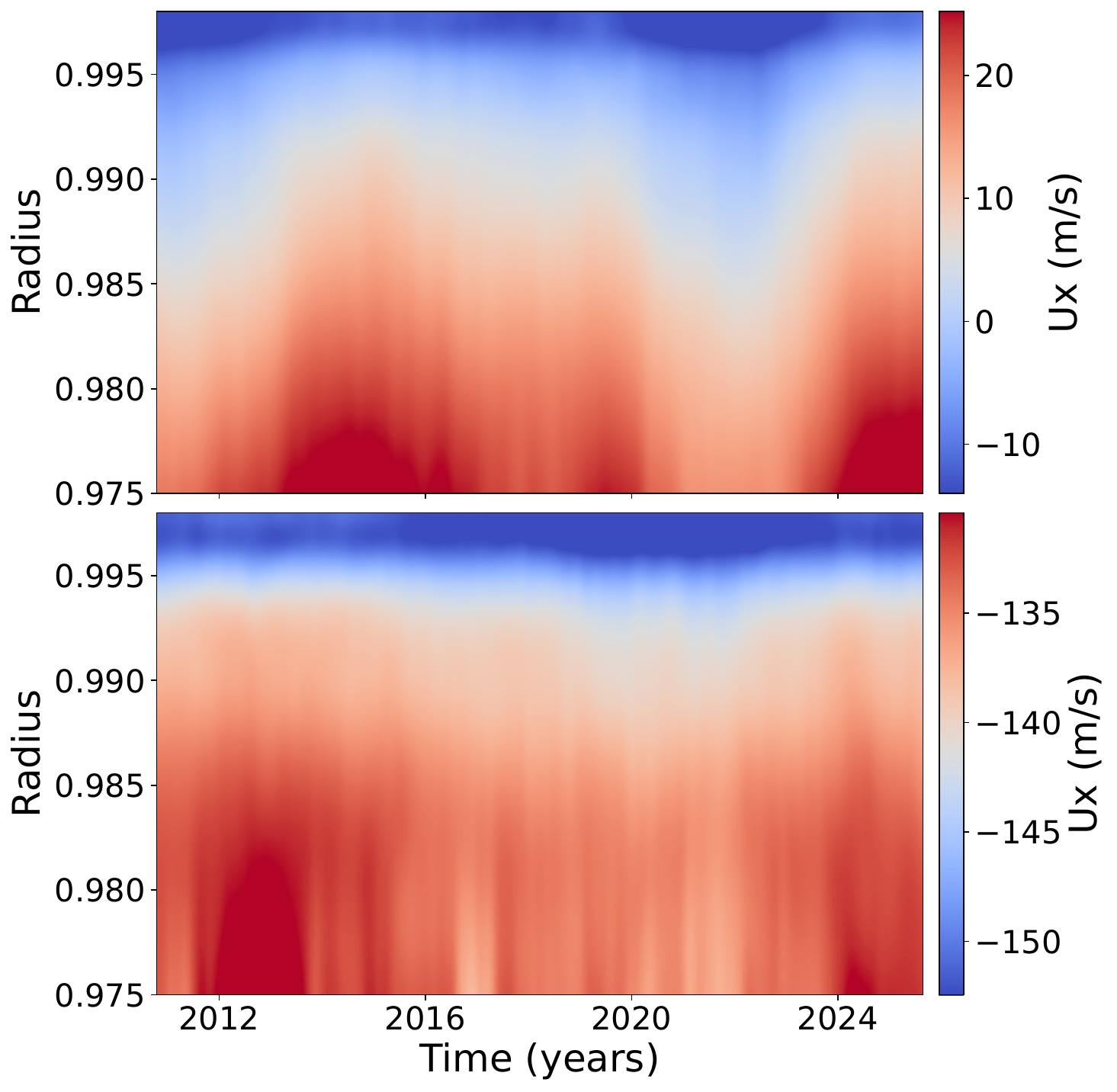}
    \caption{The north-south symmetric component of the zonal  velocity, $U_x$ plotted as a function of time and fractional radius $r/R_\odot$ at the equator (top) and at $45^\circ$ latitude (bottom). The maps were obtained with OLA; RLS results are very similar.
    }
        \label{fig:zonal}
\end{figure}

 In Fig.~\ref{fig:zonal}, we show how the rotation rate changes as a function of time and depth at different latitudes along the central meridian; we only show the component of $U_x$ that is symmetric about the equator since the antisymmetric term is small, and also because this is the component that is obtained with global-mode analysis. What is clear in  the figure is that there is a clear time-variation at each latitude.

\begin{figure}
	\includegraphics[width=0.98\linewidth]{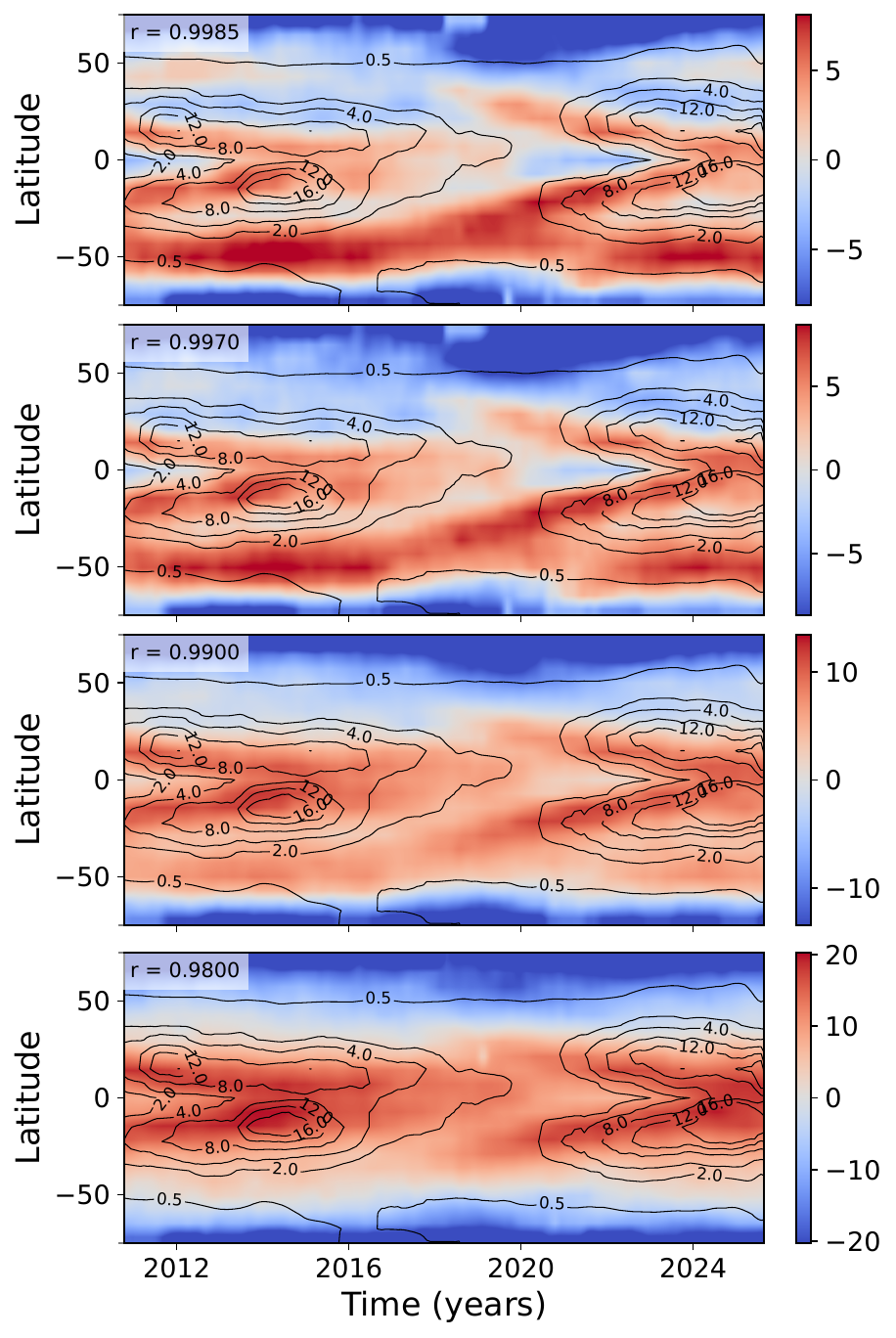}
    \caption{Zonal-velocity residuals $(\mathrm{m\,s^{-1}})$ plotted as functions of time and latitude at the fractional radii $r/R_\odot$ indicated. The residuals were obtained by first subtracting the \citet{snodgrass} differential rotation profile (Eq.~\ref{eq:snodgrass}), then subtracting the temporal and latitudinal mean at each depth to center the residuals around zero. From top to bottom, the mean values subtracted were 5.21, 6.11, 19.6, and 25.5 $(\mathrm{m\,s^{-1}})$.
   We  show OLA inversion results; RLS inversion results are very similar. Overplotted on each image are contours of the MAI.}
        \label{fig:snodgrass}
\end{figure}

In order to see the time variation better,
we first remove the \citet{snodgrass} differential rotation profile (Eq.~\ref{eq:snodgrass}), then subtract the temporal and latitudinal mean at each depth to center the residuals around zero. 
The result is shown in Fig.~\ref{fig:snodgrass}, which reveals clear equatorward-migrating bands of flow in the active latitudes. There is a significant difference between the northern and southern hemispheres, particularly in the most shallow layers. This could be a consequence of the asymmetry in the levels of magnetic activity in the two hemispheres (as noted in Section~\ref{subsec:ns}). 

Note that unlike the usual way used in global helioseismology \citep[e.g.,][etc.]{Howeetal2018, BasuAntia2019}, we have not subtracted the time-averaged flow at each latitude. Thus, in a sense, these results are more robust, since the results do not depend on the interval over which the time average was calculated. This is also why the north-south asymmetry is seen so clearly. 

\begin{figure*}
    \includegraphics[width=\linewidth]{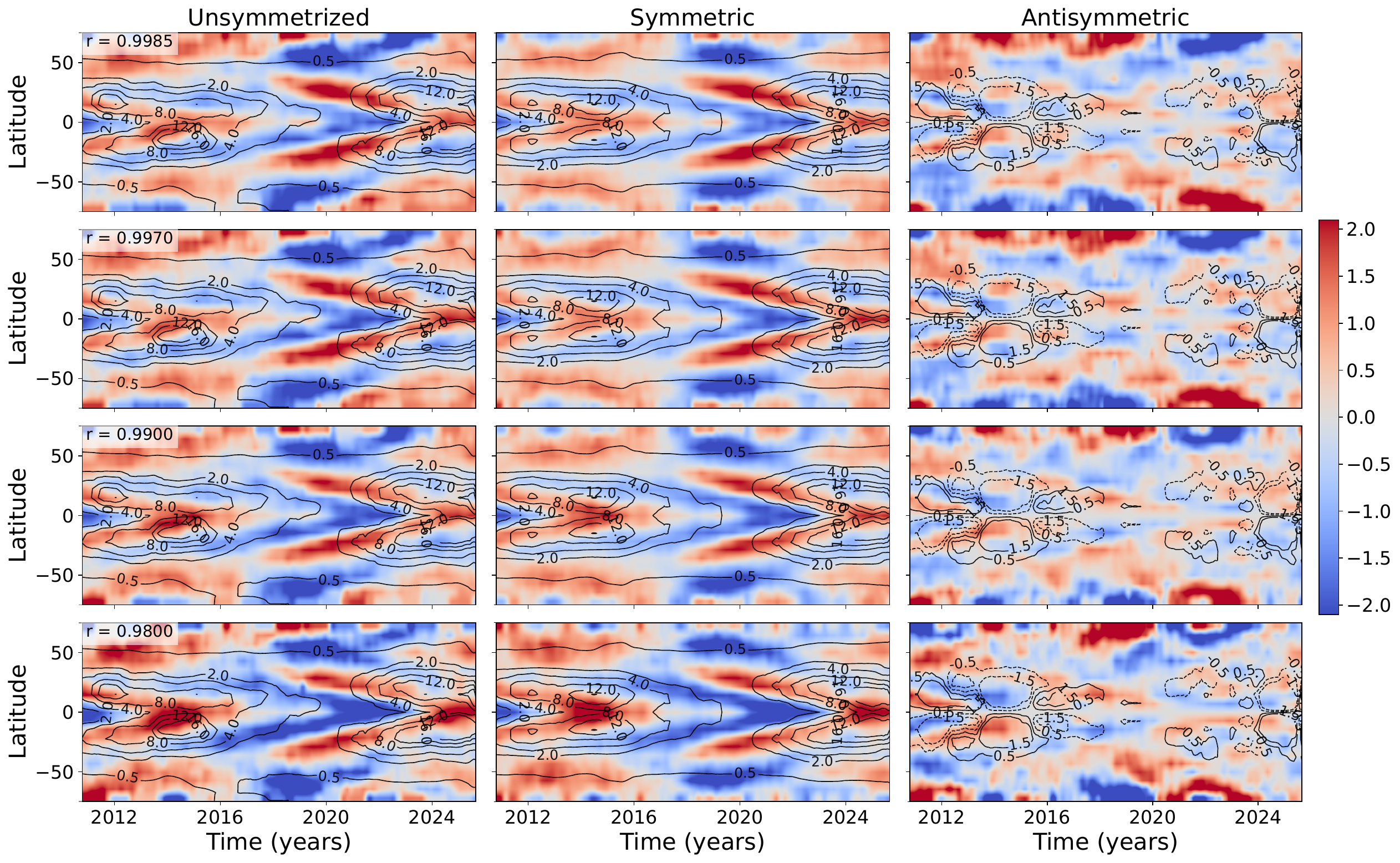}
    \caption{Zonal-velocity residuals $(\mathrm{m\,s^{-1}})$ as functions of time and latitude at the fractional radii $r/R_\odot$ indicated. 
    Left column: $\Delta U_x(r,\theta,t)$ (see Eq.~\ref{eq:resid}). Overplotted on the map are contours of the MAI.
    Middle column: North-south symmetric component $\Delta U_x(r,\theta,t)$  shown in the left column. Overplotted are contours of the north-south symmetric component of the MAI.
    Right column: The north-south antisymmetric component of $\Delta U_x(r,\theta,t)$  overplotted with the corresponding antisymmetric component of the MAI; dashed contours indicate negative values.
    The colorbar corresponds to the antisymmetric component; the total and symmetric components (left and middle columns) use a color range of $\pm 4.2\,\mathrm{m\,s^{-1}}$, twice that of the antisymmetric component ($\pm 2.1\,\mathrm{m\,s^{-1}}$).
    OLA results are shown; RLS inversions yield similar results. 
    }
    \label{fig:zonal_flow}
\end{figure*}

In order to compare our results with global mode results, we examine the flows in the same manner as in global mode analysis. i.e., we  examine the residuals of the zonal velocity that are left once the time-average of the flows is subtracted out at each depth and latitude. The residuals are calculated as 
\begin{equation}
    \Delta U_x(r,\theta,t)=U_x(r,\theta,t)-\langle U_x(r,\theta)\rangle_t, 
    \label{eq:resid}
\end{equation}
where   $\langle\;\rangle_t$ denotes a temporal average. These are shown in Fig.~\ref{fig:zonal_flow}. These migrating zonal-flow bands are often referred to as ``torsional-oscillations.'' Surface magnetic activity is quantified by the magnetic activity index (MAI; \citealt{Bogart_pipeline}), defined for each tracked tile as the unsigned HMI line-of-sight magnetic flux integrated above a 50~G threshold and then averaged in the same manner as the flow time series. The MAI contours are overplotted in Figs.~\ref{fig:snodgrass} and~\ref{fig:zonal_flow}.

\begin{figure}
    \includegraphics[width=\linewidth]{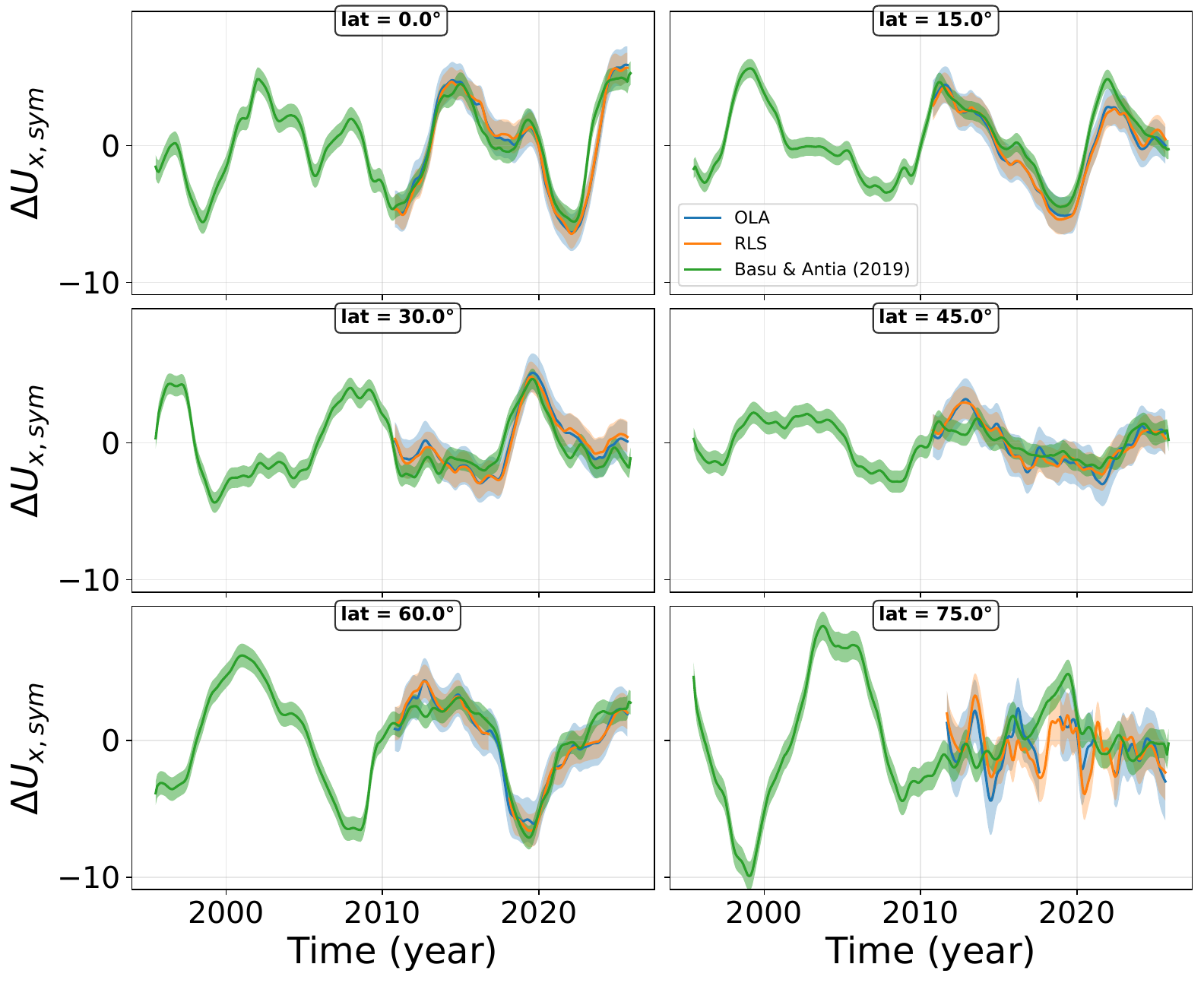}
    \caption{North–south symmetric component \(\Delta U_x(r,\theta,t)\) at \(r=0.98\,R_\odot\)  $(\mathrm{m\,s^{-1}})$ inferred from 15-degree-wide tiles using OLA (blue) and RLS (orange). The green curve shows the updated global-helioseismology results of Basu \& Antia (2019) obtained with GONG data. Global helioseismology results are courtesy of H.~M.~Antia.
    }
    \label{fig:zonal_flow_antia}
\end{figure}

Fig.~\ref{fig:zonal_flow} shows migrating bands of prograde and retrograde flows, very similar to the ones seen in global mode analysis, but we go closer to the surface. The banded flows are in the same position as seen in Fig.~\ref{fig:snodgrass}, but the retrograde flows are stronger. Even here, we find a substantial north-south asymmetry in  the flow pattern. To facilitate comparison with global mode results,  we show the symmetric component of the flows in Fig.~\ref{fig:zonal_flow} (middle column). Our results at $r=0.98\,R_\odot$ agree with those shown in Fig.~3 (left) of \citet{BasuAntia2019}. Fig.~\ref{fig:zonal_flow_antia} illustrates this agreement, comparing our symmetric component with the updated \citet{BasuAntia2019} results at several latitudes.

The north-south antisymmetric component of our results is shown in the rightmost panel of Fig.~\ref{fig:zonal_flow}. As can be seen, the antisymmetry is small in the active latitudes but is quite substantial at higher latitudes. 
There is some indication, particularly in the deepest layer ($r=0.98\,R_\odot$), that larger northern MAI coincides with larger northern $\Delta U_x$, and vice versa; in these maps, the red/blue regions align with the solid/dashed MAI contours. This correspondence is not evident at shallower depths ($r\ge 0.99\,R_\odot$).

As the MAI contours migrate equatorward from $\sim 30^\circ$ toward the equator as the cycle progresses, mimicking the movement of sunspots, the zonal flow does so too.  The MAI broadly tracks the equatorward-migrating torsional-oscillation pattern: high activity tends to be present at the boundary where the low-latitude prograde band meets the adjacent retrograde band; again, this is similar to what was seen in global mode analysis. At higher latitudes the MAI is low; however, the zonal flows show the high-latitude band that has been seen in global mode analyses \citep[e.g.,][etc.]{AntiaBasu2001, Howeetal2018}.

\subsection{Quantifying the time variations}\label{sec:cork}

\begin{figure*}
	\includegraphics[width=\linewidth]{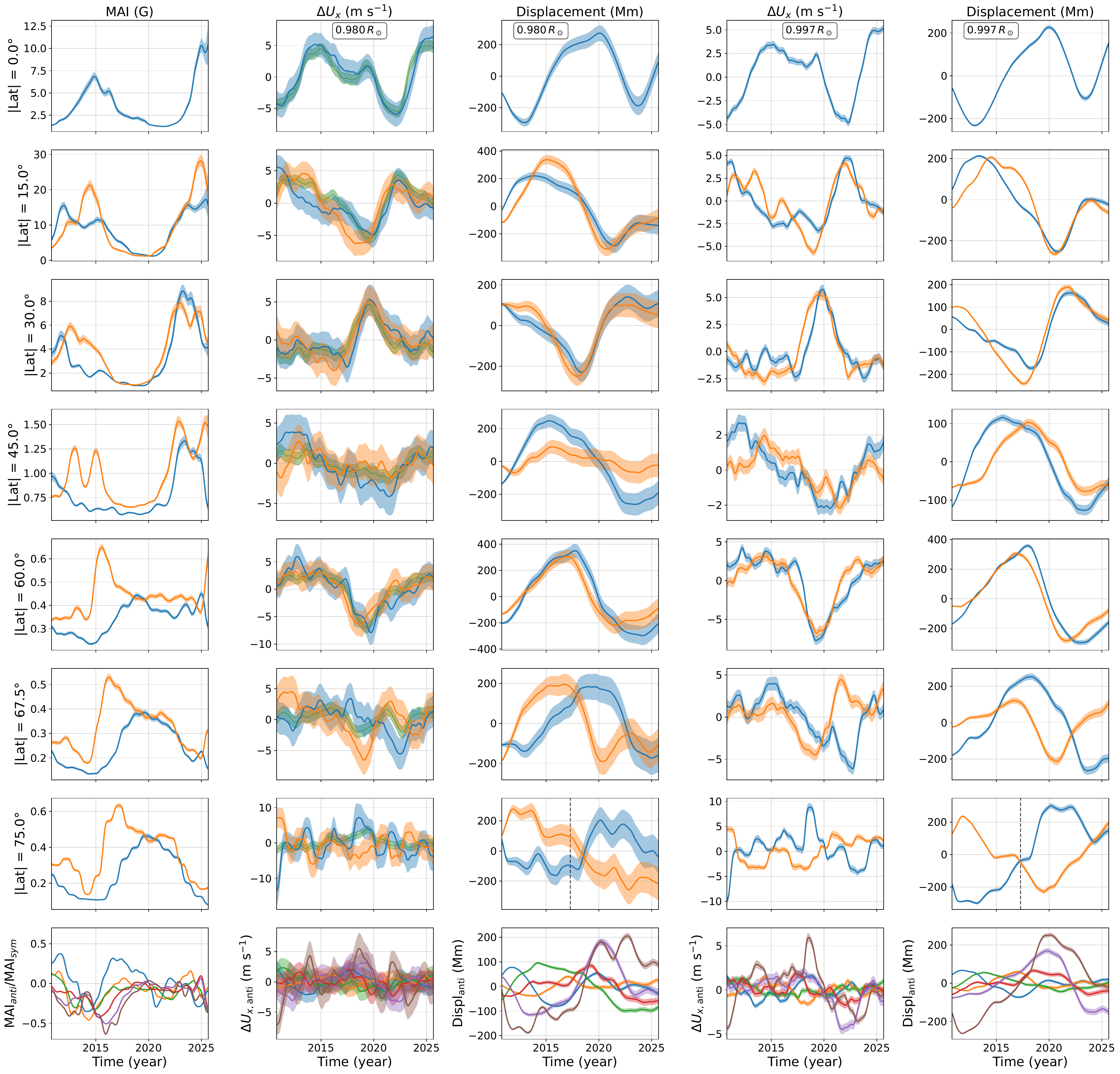}
    \caption{
    The top 7 rows show the temporal evolution of different quantities at different latitudes as noted in each row; blue denotes north and  orange  denotes  south. The first column shows the magnetic activity index, $\mathrm{MAI}(t)$. The second and fourth columns from  the left show the zonal-flow residuals, $\Delta U_x(t)$, from OLA inversions at the indicated radius. The third and fifth columns from the left show the corresponding cumulative longitudinal displacement, $\Delta s(t)$ -- Eqs.~\ref{eq:deltaphi1} and ~\ref{eq:deltaphi}. The vertical dashed line marks $t=2017.3$ yr (shown only in the $|\mathrm{lat}|=75^\circ$ panels). In all panels, the shaded regions indicate uncertainties. RLS inversions yield similar results. In the $\Delta U_x$ panels for $0.980\,R_\odot$, the green curve shows updated global-helioseismology results from \citet{BasuAntia2019} obtained with  GONG data (courtesy of H.~M.~Antia; see Fig.~\ref{fig:zonal_flow_antia}).
    Lowermost panel: the leftmost panel shows the time evolution of the fractional hemispheric asymmetry of the MAI, $\left[\mathrm{MAI}(\theta,t)-\mathrm{MAI}(-\theta,t)\right]/\left[\mathrm{MAI}(\theta,t)+\mathrm{MAI}(-\theta,t)\right]$; positive values indicate a northern excess. The second and fourth panels from the left show the antisymmetric component of the zonal-flow  residuals. The third and fifth panels from the left show the antisymmetric component of the cumulative displacement. Curves for $|\mathrm{lat}|=15^\circ, 30^\circ, 45^\circ, 60^\circ, 67\fdg5,$ and $75^\circ$ are shown in blue, orange, green, red, purple, and brown, respectively.
    }
        \label{fig:cork}
\end{figure*}

To quantify the temporal evolution of the zonal-flow bands over the solar cycle, and to enable a more direct comparison with activity diagnostics, we analyze time series of the MAI and the residual zonal velocity, $\Delta U_x$, at several representative latitudes. Additionally, we compute the cumulative longitudinal displacement implied by $\Delta U_x(t)$ by integrating $\Delta U_x$ using a constant time step of one Carrington rotation (CR),
\begin{equation}
\Delta s_{\mathrm{tot}}(t_k)= \sum_{i=1}^{k}\left[\Delta U_x(t_i)\times 10^{-6}\right]\Delta t,
\label{eq:deltaphi1}
\end{equation}
where $\Delta U_x$ is in $\mathrm{m\,s^{-1}}$, $\Delta t$ is one CR expressed in seconds, and $\Delta s_{\mathrm{tot}}$ is in Mm.
This quantity can be interpreted as the cumulative longitudinal displacement of a passive tracer advected by the residual zonal flow at fixed latitude (neglecting meridional transport).
Because $\Delta s_{\mathrm{tot}}$ is defined only up to an additive constant, we subtract its temporal mean at each latitude $\theta$ over the analysis interval, enforcing $\langle \Delta s(t)\rangle_t = 0$, i.e., we define
\begin{equation}
    \Delta s(t_k)=\Delta s_{\mathrm{tot}}(t_k)-\langle\Delta s_{\mathrm{tot}}(t)\rangle_t,
    \label{eq:deltaphi}
\end{equation}
where the angular brackets denote the temporal mean.
We have verified that using the full time span or restricting to an $\approx 11$-yr interval yields consistent results. Uncertainties in $\Delta s$ increase with time as the uncertainties in $\Delta U_x$ accumulate through the integration. These results are shown in Fig.~\ref{fig:cork}. From left to right, the panels show the time variation of the MAI, the residual zonal velocity $\Delta U_x$ at $0.98\,R_\odot$, the displacement $\Delta s$ at $0.98\,R_\odot$, $\Delta U_x$ at $0.997\,R_\odot$, and $\Delta s$ at $0.997\,R_\odot$.

As can be seen from Fig.~\ref{fig:cork}, a north-south asymmetry in the MAI is evident at all latitudes. Although the MAI amplitude decreases strongly with latitude, it exhibits coherent temporal variations that exceed the noise level and thus provide a meaningful proxy for the evolution of the fields, even at high latitudes. No significant north-south asymmetry is visible in the $U_x$ residuals in the low and mid latitudes, particularly in the deeper layer at $0.98\,R_\odot$, although there are hints of north-south asymmetry even at the low and mid latitudes in the shallower layers. 
However, deviations from hemispherical symmetry become apparent at high latitudes.
For $|\theta|\ge 67.5^\circ$, the hemispheric differences are significant and the values of $\Delta U_x$  are larger than those at lower latitudes. 
The cumulative longitudinal displacement $\Delta s$ also shows larger amplitudes at high latitudes and a pronounced hemispheric asymmetry; at $|\theta|=75^\circ$ it is nearly antisymmetric, with large displacements of opposite signs in the two hemispheres. The displacements at latitudes $\pm 75^\circ$ intersect near zero and reverse sign around 2017.3 in the shallower layer (marked by the vertical dashed line), during the descending phase of cycle 24, and about one year later (2018.3), closer to solar minimum, in the deeper layer.
Overall, the high-latitude zonal-flow displacement appears to follow the evolution of the solar cycle.  

In October 2018, a change in the HMI focus control produced a very small (less than 0.1\%) plate-scale change. In Paper~II we show that this change corresponds to a slight ($\sim$0.2~Mm) shift in the inferred location of the near-surface enhanced-shear layer in the dimensionless radial shear $\partial\ln\Omega/\partial\ln r$. We find no corresponding signature in the zonal flows or in the displacement $\Delta s$ analyzed here, and the north–south convergence at $|\theta|=75^\circ$ occurs well before the 2018 focus change (vertical dashed line in Fig.~\ref{fig:cork}).

The lowermost panels of Fig.~\ref{fig:cork} show the antisymmetric components of all three observables. The magnetic-field antisymmetry is weaker in Cycle~25 than in Cycle~24. The southern hemisphere dominates, particularly during $\approx 2012$--2018. In contrast, the zonal-flow residuals are consistent with a symmetric distribution within the uncertainties. As already indicated by the upper panels of  Fig.~\ref{fig:cork}, $\Delta s$ shows a large antisymmetric component at high latitudes.
This high-latitude asymmetry in $\Delta s$ persists despite the fact that the residual definition (Equation~\ref{eq:resid}) removes any time-independent north–south antisymmetry associated with line-of-sight effects (Sections~\ref{subsec:ns} and \ref{subsec:EW}).

Examining $\Delta s$ in the $75^\circ$ latitude band (Fig.~\ref{fig:cork}), we find large positive values in the south early in cycle 24 and negative values in the north; early in cycle 25, the sense reverses.
Interpreted as the cumulative longitudinal displacement of a passive tracer (e.g., a magnetic element treated as being advected by the flow), this corresponds in the shallowest layer to strong cumulative prograde transport in the south during $\approx$2011-2014, followed by a corresponding episode in the north during $\approx$2019-2023.
Such longitudinal transport may influence the redistribution of magnetic elements at high latitudes and thereby potentially contribute to the evolution and reversal of the polar field.
Motivated by this possibility, we compare our results at $75^\circ$ latitude with the mean polar field, defined as the radial magnetic-field strength averaged above \(60^\circ\) latitude and measured by HMI (JSOC data series \texttt{hmi.meanpf\_720s}), following \citet[][Fig.~2c]{Sun2015} and using the updated version maintained by \citet{Sun_and_Bobra}. 
The unsigned mean polar field closely tracks the MAI at $75^\circ$ (Fig.~\ref{fig:cork}, left panel, second-to-last row).
The larger displacement in the south during $\approx$2011-2014 may be associated with the more rapid polar-field reversal in the southern hemisphere than in the north around 2014. A similar correspondence is not evident during the 2024 polar-field reversal: the larger displacement in the north is not accompanied by a comparably more rapid northern reversal, but only by a slight increase in the reversal rate relative to the south. 
Rather, the strongest positive antisymmetry in $\Delta s$ ($\approx$2019–2023) coincides with a weakening of the southern polar field while the northern polar field reaches a maximum and then remains approximately constant.
We note that these intervals of large positive $\Delta s$ occur at different phases of the solar cycle: the southern episode ($\approx$2011-2014) falls in the late rising phase of cycle 24, approaching solar maximum, whereas the northern episode ($\approx$2019–2023) occurs in the early rising phase of cycle 25, following solar minimum. 
This difference in timing relative to the cycle may help explain why the apparent correspondence with the polar-field evolution is stronger in cycle 24 than in cycle 25.

\begin{figure*}
    \includegraphics[width=\linewidth]{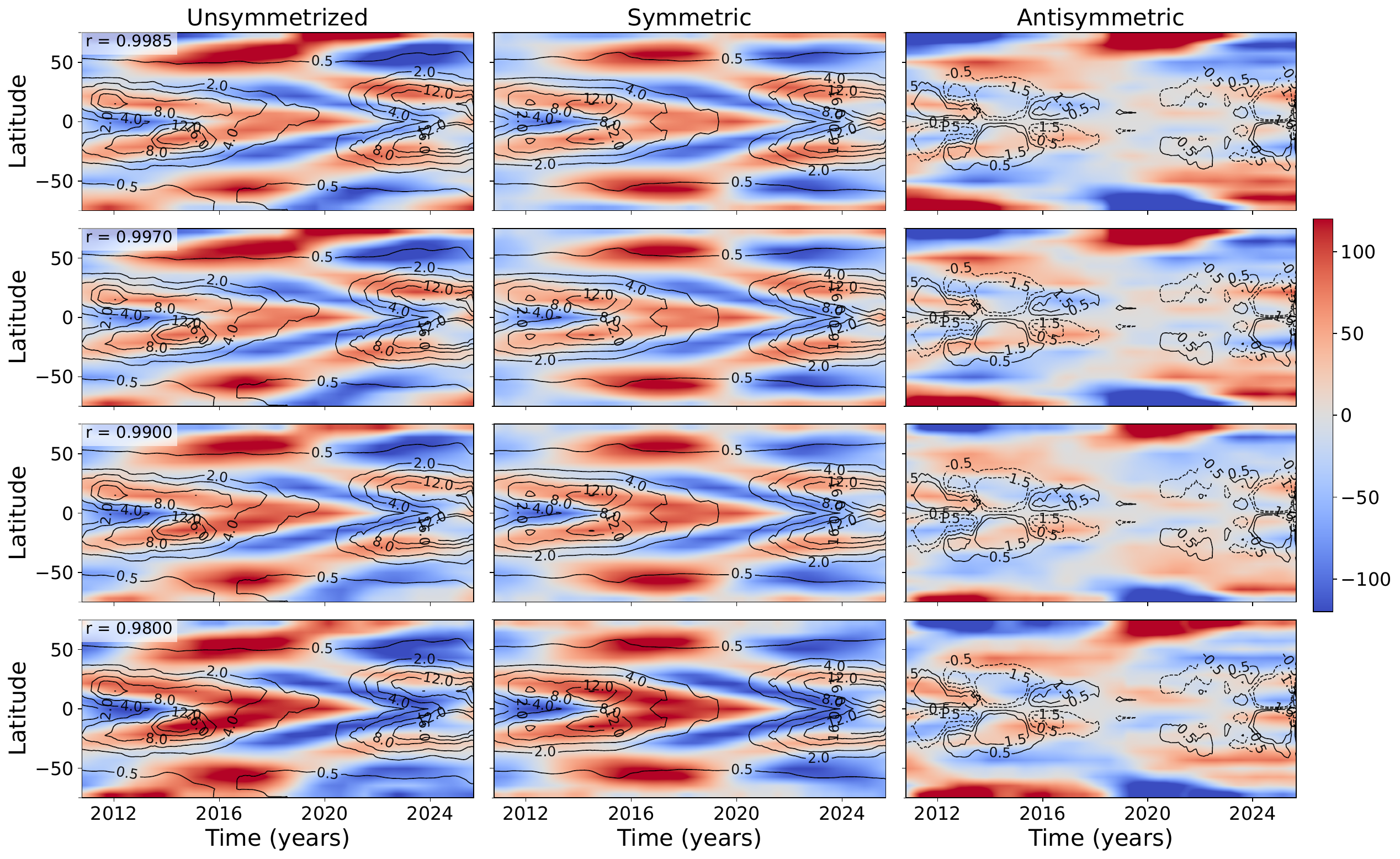}
    \caption{
    Same format as Fig.~\ref{fig:zonal_flow}, but for the cumulative longitudinal displacement $\Delta s(r,\theta,t)$ (Mm) instead of the zonal-velocity residual $\Delta U_x$. Contours show MAI (left), its north–south symmetric component (middle), and antisymmetric component (right; dashed negative). The colorbar corresponds to the antisymmetric component; the total and symmetric components (left and middle columns) are shown with a color range of $\pm 240\,\mathrm{Mm}$, while the antisymmetric component (right column) uses $\pm 120\,\mathrm{Mm}$. OLA results are shown; RLS is similar.
    }
    \label{fig:disp}
\end{figure*}

Figure~\ref{fig:disp} shows the displacement in the same layout as the torsional oscillations in Figure~\ref{fig:zonal_flow}.
%
%
The displacement lags the zonal-flow residuals by $2.7$~yr. The north--south symmetric component yields slightly higher peak correlations than the unsymmetrized signal, with $\langle r_{\rm best}\rangle \approx 0.81$. At zero lag the correlation is consistent with zero, indicating that the agreement is dominated by the applied time shift. This behavior is consistent, to first approximation, with a sinusoidal torsional oscillation and a displacement in temporal quadrature (a $90^\circ$phase offset for a period of$\sim 11$ yr).
Since $\Delta s$ is the time integral of the residual flows, it suppresses short-term variability.
In Figure~\ref{fig:disp}, $\Delta s(\theta,t)$ also shows the migratory pattern more clearly than the corresponding $\Delta U_x(\theta,t)$ maps. Near $|\theta|\simeq 58^\circ$, $\Delta s$ exhibits alternating bands of positive and negative displacement that remain nearly stationary in latitude, persisting for roughly half a cycle. These features are approximately north--south symmetric.
Toward the end of each half-cycle, the bands develop equatorward-moving branches; in the north they additionally bifurcate, producing a separate high-latitude branch near $\sim 75^\circ$ (the observational limit of our analysis), which is not seen in the south.
At higher latitudes (\(|\theta|\gtrsim 60^\circ\)), \(\Delta s\) is strongly north–south antisymmetric, reversing sign between hemispheres from about one half-cycle to the next, consistent with the \(\Delta s(t)\) time series in Figure~\ref{fig:cork} (second-to-last row; third and fifth columns).

To quantify the correlation and relative timing between the magnetic and flow signals suggested by Fig.~\ref{fig:cork}, we perform a lagged-correlation analysis between MAI and the zonal-flow related quantities at each latitude. For each latitude at $r=0.98$, 0.99, 0.997, and \(0.9985\,R_\odot\) (the same radii used in Section~\ref{subsec:zonal}), we analyze time series ($N=200$) of MAI and either the zonal-flow residuals $\Delta U_x$ or the cumulative longitudinal displacement $\Delta s$. 
After standardizing each series by subtracting its time mean and dividing by its standard deviation, we compute the Pearson correlation at zero lag, \(r_0\), and the cross-correlation $r(\tau)$ for $\tau\in[-99,+99]$~CR (requiring $\ge 60$ overlapping samples). We identify the lag $\tau_{\max}$ that maximizes $|r(\tau)|$. 
Significance is estimated from 5000 phase-randomized surrogates that preserve each series’ power spectrum (and hence autocorrelation), yielding two-sided p-values \(p_{r0}\) for \(r_0\) and a global p-value \(p_{\max}\) for the lag-scanned maximum. 
We report \(r_0\), \(p_{r0}\), \(\tau_{\max}\), \(r(\tau_{\max})\), and \(p_{\max}\) for \(\Delta s\) in Table~\ref{tab:lagcorr_098} ($r=0.98\,R_\odot$); the corresponding results for the other radii are provided in Appendix~\ref{app:lagcorr_deep} (Tables~\ref{tab:lagcorr_099}, \ref{tab:lagcorr_0997}, and \ref{tab:lagcorr_09985}). At all four radii, we find that the MAI--$\Delta U_x$ correlations are generally not statistically significant.
Because $\Delta s$ is the time integral of $\Delta U_x$, MAI–$\Delta s$ correlations are more often statistically significant than MAI–$\Delta U_x$ correlations.

Table~\ref{tab:taumax_vs_r} summarizes the MAI--\(\Delta s\) lag results across all four radii (and both inversion methods). Statistically significant and OLA/RLS-consistent correlations occur at only a subset of latitudes.
There is a consistent negative lag at \(15^\circ\) latitude, indicating that \(\Delta s\) leads MAI. The results are robust to the inversion method (OLA and RLS agree closely) and are statistically significant: at \(15^\circ\)S, \(\tau_{\max}\approx-4.2\) yr in the deepest layer, increasing in magnitude to \(\approx-4.6\) yr in the shallowest layer; at \(15^\circ\)N, \(|\tau_{\max}|\) is larger, with \(\tau_{\max}\approx-4.7\) yr in the deepest layer increasing in absolute value to \(\approx-6.5\) yr in the shallowest layer. This latitude lies within the activity belts, where sunspots are concentrated near the shear boundary between the equatorward-migrating prograde ($\Delta U_x>0$) band and the adjacent retrograde ($\Delta U_x<0$) band (Section~\ref{subsec:zonal}).
At intermediate latitudes (\(30^\circ\)–\(45^\circ\)) and at the equator, the lags are poorly constrained (\(p>0.10\) or \(|\tau_{\max}|\gtrsim 7\) yr, close to the maximum lag searched) and show poor OLA/RLS agreement. An exception is \(30^\circ\)S, where correlations are significant at \(p<0.10\) (with OLA/RLS agreement except at \(0.98\,R_\odot\)); notably, the preferred lag changes sign with radius, from \(\tau_{\max}\approx+4.3\)–\(4.8\) yr at the two deeper layers, transitioning to \(\tau_{\max}\approx-1.5\) yr in the two shallower layers.
At southern high latitudes (\(60^\circ\)S and \(67\fdg5\)S), the lags are poorly constrained (\(p>0.10\) or \(|\tau_{\max}|\gtrsim 7\) yr) and show poor OLA/RLS agreement. In the north, by contrast, the lags are well constrained, with good OLA/RLS agreement: at \(60^\circ\)N, \(\tau_{\max}\approx+3.5\) yr at all radii, while at \(67\fdg5\)N, \(\tau_{\max}\approx+5.6\) yr in the deepest layer, decreasing to \(\approx+4.6\) yr in the shallower layers.
At \(75^\circ\) latitude, the situation reverses: \(75^\circ\)N is poorly constrained, whereas \(75^\circ\)S is well constrained (\(p<0.10\)) and shows good OLA/RLS agreement. At \(75^\circ\)S, \(\tau_{\max}\approx-4.9\) yr in the two deepest layers, increasing to \(\approx-5.5\) yr in the shallower layers.
At \(75^\circ\)S, the preferred lag is about half a solar cycle, so \(\Delta s\) is roughly in antiphase with the unsigned MAI. If we instead compare with the signed polar field, the south polar field was negative (and the north polar field positive) throughout 2014--2024, representing more than 70\% of our data interval. This implies that \(\Delta s\) is approximately in phase with the polar-field polarity.
Although the lag at \(75^\circ\)N is not well constrained, the best-fit value is near zero, suggesting an in-phase relationship.

Overall, the inferred lag shows a radial dependence with small amplitude ($\lesssim 1$ yr), but the radial trend varies by latitude. $|\tau_{\max}|$ is approximately constant with radius at $60^\circ$N, increases toward the surface at $15^\circ$N, $15^\circ$S, and $75^\circ$S, and decreases toward the surface at $67\fdg5$N. At $30^\circ$S, $\tau_{\max}$ reverses sign, from positive at depth to negative closer to the surface, with a much larger net change.
We see a clear hemispheric asymmetry at high latitudes (\(\ge60^\circ\)): \(\tau_{\max}\) is positive at \(60^\circ\)N and \(67\fdg5\)N (but not well determined at \(60^\circ\)S and \(67\fdg5\)S), whereas it is negative at \(75^\circ\)S (with \(75^\circ\)N not well determined).
These results raise the question of whether the inferred lag asymmetry reflects hemispheric differences in the timing of cycle evolution, potentially due to differences in the coupling between magnetic-field evolution and large-scale flows at mid to high latitudes, and whether this behavior varies from one cycle to the next or with phase. Such variability could contribute to the limited number of significant detections in an analysis that spans multiple phases and cycles.

\begin{deluxetable}{lrrrrrr}
\tablewidth{1.0\columnwidth} 
\tablecaption{Cross-correlation summary for the cumulative longitudinal displacement $\Delta s$ at radius $0.98\,R_\odot$ \label{tab:lagcorr_098}}
\tablehead{
\colhead{Inv} &
\colhead{lat} &
\colhead{$r_0$} &
\colhead{$p_{r0}$} &
\colhead{$\tau_{\max}$} &
\colhead{$r(\tau_{\max})$} &
\colhead{$p_{\max}$}}
\tabletypesize{\scriptsize}
\startdata
OLA &   0.0 & -0.218 & 0.6447 &  3.923           &  0.796 & 0.2230 \\
RLS &   0.0 & -0.231 & 0.6137 & -7.615$\dagger$  &  0.804 & 0.2014 \\
OLA &  15.0 &  0.196 & 0.8492 & -4.846           & -0.903 & 0.2925 \\
RLS &  15.0 &  0.163 & 0.8776 & -4.615           & -0.901 & 0.3439 \\
OLA & -15.0 &  0.190 & 0.8110 & -4.154           & -0.945 & 0.0092 \\
RLS & -15.0 &  0.195 & 0.8014 & -4.077           & -0.938 & 0.0138 \\
OLA &  30.0 &  0.744 & 0.1676 & -5.692           & -0.914 & 0.4055 \\
RLS &  30.0 &  0.724 & 0.1946 &  7.154$\dagger$    & -0.919 & 0.3627 \\
OLA & -30.0 &  0.782 & 0.1228 & -6.000           & -0.967 & 0.0314 \\
RLS & -30.0 &  0.795 & 0.1176 &  4.846           & -0.960 & 0.0648 \\
OLA &  45.0 & -0.772 & 0.1522 &  7.615$\dagger$    &  0.912 & 0.2987 \\
RLS &  45.0 & -0.742 & 0.2214 &  7.615$\dagger$    &  0.933 & 0.1562 \\
OLA & -45.0 & -0.465 & 0.4283 & -7.615$\dagger$    &  0.858 & 0.1262 \\
RLS & -45.0 & -0.503 & 0.4153 & -7.615$\dagger$    &  0.857 & 0.1394 \\
OLA &  60.0 & -0.185 & 0.8550 &  3.538           & -0.945 & 0.0694 \\
RLS &  60.0 & -0.162 & 0.8736 &  3.538           & -0.946 & 0.0582 \\
OLA & -60.0 &  0.468 & 0.1984 &  5.538           & -0.720 & 0.4401 \\
RLS & -60.0 &  0.085 & 0.8690 &  2.846           &  0.790 & 0.0928 \\
OLA &  67.5 &  0.805 & 0.2685 &  5.615           & -0.988 & 0.0054 \\
RLS &  67.5 &  0.688 & 0.4137 &  5.462           & -0.987 & 0.0004 \\
OLA & -67.5 &  0.329 & 0.6931 &  3.769           & -0.899 & 0.3825 \\
RLS & -67.5 &  0.675 & 0.0418 & -7.538$\dagger$    & -0.948 & 0.0030 \\
OLA &  75.0 &  0.717 & 0.2258 &  1.385           &  0.903 & 0.4123 \\
RLS &  75.0 &  0.685 & 0.2667 &  1.231           &  0.777 & 0.9014 \\
OLA & -75.0 & -0.065 & 0.9462 & -5.000           &  0.957 & 0.0018 \\
RLS & -75.0 & -0.048 & 0.9660 & -4.846           &  0.915 & 0.0604 \\
\enddata
\tablecomments{
Column Inv lists the inversion method, $\tau_{\max}$ is in years, and $p_{\max}$ accounts for the scanned lag window $\tau\in[-99,99]$ CR. Positive lags (\(\tau>0\)) indicate that \(\Delta s\) lags MAI by \(\tau\).
$\dagger$\ marks $|\tau_{\max}|>90$ CR.
}
\end{deluxetable}

\begin{deluxetable}{lcccc}
\tablecaption{Lags $\tau_{\max}$ (years) maximizing $|r(\tau)|$ for MAI--$\Delta s$ cross-correlations at four radii.\label{tab:taumax_vs_r}}
\tablehead{
\colhead{lat} &
\colhead{$0.98\,R_\odot$} &
\colhead{$0.99\,R_\odot$} &
\colhead{$0.997\,R_\odot$} &
\colhead{$0.9985\,R_\odot$}
}
\startdata
\phantom{-}75.0 & (0.0) & (0.0) & -7.5/(0.0) & -7.4/-7.5 \\
\phantom{-}67.5 & 5.6/5.5 & (-7.6)/(4.2) & 4.5/4.6 & 4.5* \\
\phantom{-}60.0 & 3.5* & 3.5 & 3.6/3.5 & 3.6/3.5 \\
\phantom{-}45.0 & (0.0)/(7.6) & (0.0) & (0.0) & (0.0) \\
\phantom{-}30.0 & (0.0) & (0.0) & (0.0) & (0.0) \\
\phantom{-}15.0 & (-4.8)/(-4.6) & -5.9*/(-6.0) & -6.5/-6.4 & -6.4/-6.5 \\
\phantom{-}0.0 & (3.9)/(-7.6) & (-7.6) & (-7.6) & -7.6* \\
-15.0 & -4.2/-4.1 & -4.2 & -4.5/-4.4 & -4.5/-4.6 \\
-30.0 & -6.0/4.8* & 4.3* & -1.5* & -1.6*/-1.5 \\
-45.0 & (-7.6) & 0.0/-7.1 & -7.1/-7.3 & -7.1/-7.4 \\
-60.0 & (0.0)/2.8* & (0.0)/(5.9) & (5.8)/(5.7) & (5.6)/5.5* \\
-67.5 & (3.8)/-7.5 & 3.7/3.8 & -7.6* & (3.5)/-7.6 \\
-75.0 & -5.0/-4.8* & -5.0/-4.8* & -5.4* & (-5.4)/-5.5 \\
\enddata
\tablecomments{
Entries columns $\tau_{\max}$ in years as OLA/RLS (rounded to two significant digits); a single value indicates identical OLA and RLS results. For each method, $\tau_{\max}$ is reported only when the lag-scanned correlation is more significant than the zero-lag correlation ($p_{\max}<p_{r0}$); otherwise the entry is set to $0.0$ (i.e., lag 0 is favored). Positive lags ($\tau_{\max}>0$) indicate that $\Delta s$ lags MAI by $\tau_{\max}$. 
Significance flags are applied to each value: parentheses denote \(p_{\max}\ge 0.10\); * denotes \(0.05 \le p_{\max} < 0.10\); unmarked values have \(p_{\max}<0.05\). Lags with \(|\tau_{\max}|>7\) yr occur at the edge of the searched lag range.}
\end{deluxetable}

\section{Discussion and Conclusions}
\label{sec:conclusion}

We have done a helioseismic study of the near-surface shear layer using the local helioseismic technique of ring-diagram analysis using HMI data. This has allowed us to study the rotation rate and their changes from a depth of 1~Mm to about 17~Mm below the solar surface. 

We find that there is a  north-south asymmetry in the time-averaged flows. However, we cannot be certain as to whether this is a long-term feature of solar rotation, or whether it is because of the asymmetry in solar activity over the period for which HMI data are available. The north-south difference is only about 1\% of the rotation rate, and we  cannot completely discount other systematic effects in the data that could cause this effect. 

The time variation of the NSSL echoes that of the deeper layers and shows the ``torsional oscillation'' signal.  However, we find that the time variation can be observed even without subtracting the mean rotation rate; all that needs to be done is to remove the Snodgrass differential-rotation profile. This results in a pattern that is robust to issues concerning the time span of the data. 

We have also followed what is done in global-mode analyses, i.e., subtract the time-averaged rotation rate at each depth and latitude from the rotation rate of each epoch. We see the torsional-oscillation pattern even in a layer that is as shallow as 1~Mm. Unlike other local helioseismic investigations of the NSSL \citep[e.g.,][etc.]{Zhao2014, komm2022, Komm2025} we have not needed to average the results over a large depth range to determine the time variations. In deeper layers, the pattern agrees with that found by global mode analyses. 

We have defined a cumulative longitudinal displacement, $\Delta s$, implied by the zonal-flow residuals $\Delta U_x$, and find that it exhibits substantial north--south asymmetry at high latitudes, with variability on timescales comparable to the solar cycle. The displacements at $75^\circ$ latitude are largely antisymmetric and exhibit a depth-dependent timing of the sign reversal: the transition occurs around 2017.3 near the surface but approximately one year later at $0.98\,R_\odot$, closer to solar minimum.

As a kinematic diagnostic,
$\Delta s$ may be interpreted as the displacement of a passive tracer, such as a magnetic element advected by the residual zonal flow at fixed latitude, neglecting meridional circulation.
Interpreted in this way, the evolution of $\Delta s$ appears to show some temporal correspondences with that of the polar field: the enhanced southern displacement at $75^\circ$ during $\approx$2011-2014 occurs near the epoch of polar-field reversal, during which the southern hemisphere reversed substantially more rapidly than the north. By contrast, the enhanced northern displacement that developed during $\approx$2019–2023 coincides with a period during which the northern polar field remained near its maximum strength while the southern polar field weakened.
These temporal correspondences suggest that longitudinal advection associated with residual zonal flows may contribute to magnetic-flux redistribution at high latitudes, alongside the dominant poleward transport responsible for polar-field evolution.
We note that the strongest positive \(\Delta s\) episodes occur at different phases of the activity cycle, which may help explain why the apparent correspondence with polar-field evolution differs between the two cycles.

To explore the relationship between magnetic activity (MAI) and the flow signals,
motivated in particular by the high-latitude behavior, we performed a lagged-correlation analysis at each latitude.
We do not find a significant correlation between the MAI and the velocity residuals $\Delta U_x$. In contrast, the cumulative longitudinal displacement shows statistically significant correlations with MAI at a subset of latitudes. Among the statistically significant cases ($p<0.10$) for which OLA and RLS give consistent results, the best-fit lag $\tau_{\max}$ spans $\sim\pm6$ yr across latitudes and radii.
The variation of the lag with depth is generally weak (\(\lesssim 1\) yr) and shows different trends, ranging from increases toward the surface (\(15^\circ\)N, \(15^\circ\)S, and \(75^\circ\)S) to decreases (\(67\fdg5\)N) or near constancy (\(60^\circ\)N). At $30^\circ$S, $\tau_{\max}$ changes by $\sim 6$ yr, from $\approx+4.5$ yr in the deepest layer to $\approx-1.5$ yr in the shallowest layer.
Hemispheric asymmetry is evident at high latitudes ($\ge 60^\circ$): the statistically significant lags are positive in the north, with $\tau_{\max}\simeq +3.5$ yr at $60^\circ$N and $\tau_{\max}\approx +5$ yr at $67\fdg5$N, and negative in the south, with $\tau_{\max}\approx -5$ yr at $75^\circ$S; other cases are not well constrained.
The small set of latitudes with robust lag determinations 
may indicate that the coupling is more complex. The 
north–south differences in \(\tau_{\max}\) may be associated with hemispheric offsets in cycle progression, potentially reflecting hemispheric variations in the coupling between magnetic evolution and the large-scale flows at mid to high latitudes. It will be useful to test whether the inferred lags persist from one cycle to the next and whether they depend on cycle phase.

\begin{acknowledgments}
We thank H.~M.~Antia for providing us with the updated global-mode results. The authors acknowledge support from NASA grant 80NSSC25K7669. This research was also supported in part by NASA Contract NAS5-02139 to Stanford University and by the COFFIES DSC Cooperative Agreement 80NSSC22M0162. This work uses data from the Helioseismic and Magnetic Imager. HMI data are courtesy of NASA/SDO and the HMI science team. The data used in this article are publicly available from the Joint Science Operations Center at jsoc.stanford.edu.
\end{acknowledgments}

\facilities{HMI(SDO), JSOC(Stanford and Lockheed)}

\appendix
\section{Lagged-correlation results at different radii}
\label{app:lagcorr_deep}

Table~\ref{tab:lagcorr_098} summarizes the lagged-correlation analysis between MAI and the
cumulative longitudinal displacement $\Delta s$ at $r=0.98\,R_\odot$.
For completeness, we report the corresponding results at shallower radii,
 0.99, 0.997, and $0.9985\,R_\odot$, in Tables~\ref{tab:lagcorr_099}, \ref{tab:lagcorr_0997}, and \ref{tab:lagcorr_09985}, respectively, as in Section~\ref{subsec:zonal}.
The analysis procedure is identical to that
described in Section~\ref{sec:cork}: each Carrington-rotation time series is
standardized, cross-correlations are evaluated over lags
$\tau\in[-99,+99]$~CR with a minimum overlap of 60 samples, and
significance is assessed using 5000 phase-randomized surrogates. As in the main text,
we tabulate 
\(r_0\), \(p_{r0}\), the lag \(\tau_{\max}\) maximizing \(r(\tau_{\max})\), $r(\tau_{\max})$, and the corresponding global significance $p_{\max}$.

\begin{deluxetable}{lrrrrrr}
\tablecaption{Cross-correlation summary for the cumulative longitudinal displacement $\Delta s$ at a radius of $0.99\,R_\odot$ \label{tab:lagcorr_099}}
\tablehead{
\colhead{Inv} &
\colhead{lat} &
\colhead{$r_0$} &
\colhead{$p_{r0}$} &
\colhead{$\tau_{\max}$} &
\colhead{$r(\tau_{\max})$} &
\colhead{$p_{\max}$}}
\tabletypesize{\scriptsize}
\startdata
OLA &   0.0 & -0.182 & 0.6971 & -7.615$\dagger$  &  0.817 & 0.1516 \\
RLS &   0.0 & -0.194 & 0.6649 & -7.615$\dagger$  &  0.827 & 0.1300 \\
OLA &  15.0 &  0.422 & 0.6627 & -5.923           & -0.935 & 0.0752 \\
RLS &  15.0 &  0.450 & 0.6389 & -6.000           & -0.931 & 0.1028 \\
OLA & -15.0 &  0.267 & 0.7089 & -4.154           & -0.958 & 0.0056 \\
RLS & -15.0 &  0.261 & 0.7133 & -4.154           & -0.956 & 0.0058 \\
OLA &  30.0 &  0.764 & 0.1388 & -1.000           &  0.877 & 0.6391 \\
RLS &  30.0 &  0.758 & 0.1452 &  7.308$\dagger$    & -0.881 & 0.6033 \\
OLA & -30.0 &  0.687 & 0.2164 &  4.308           & -0.961 & 0.0542 \\
RLS & -30.0 &  0.675 & 0.2362 &  4.308           & -0.958 & 0.0768 \\
OLA &  45.0 & -0.740 & 0.2545 &  7.615$\dagger$    &  0.916 & 0.2703 \\
RLS &  45.0 & -0.753 & 0.2006 &  7.615$\dagger$    &  0.913 & 0.2869 \\
OLA & -45.0 & -0.727 & 0.0064 & -7.231$\dagger$    &  0.916 & 0.0212 \\
RLS & -45.0 & -0.717 & 0.0284 & -7.077$\dagger$    &  0.927 & 0.0122 \\
OLA &  60.0 & -0.200 & 0.8390 &  3.462           & -0.964 & 0.0114 \\
RLS &  60.0 & -0.128 & 0.9004 &  3.462           & -0.970 & 0.0046 \\
OLA & -60.0 &  0.446 & 0.2314 &  5.846           & -0.739 & 0.3663 \\
RLS & -60.0 &  0.332 & 0.4865 &  5.923           & -0.729 & 0.3867 \\
OLA &  67.5 &  0.273 & 0.8048 & -7.615$\dagger$    & -0.963 & 0.1656 \\
RLS &  67.5 &  0.294 & 0.7952 &  4.154           & -0.969 & 0.2250 \\
OLA & -67.5 &  0.270 & 0.6619 &  3.692           & -0.952 & 0.0150 \\
RLS & -67.5 &  0.221 & 0.7814 &  3.769           & -0.955 &  0.0276 \\
OLA &  75.0 &  0.673 & 0.4271 & -7.077$\dagger$    & -0.900 & 0.5253 \\
RLS &  75.0 &  0.675 & 0.4123 & -7.154$\dagger$    & -0.873 & 0.7598 \\
OLA & -75.0 & -0.354 & 0.6615 & -5.000           &  0.967 & 0.0140 \\
RLS & -75.0 & -0.242 & 0.7546 & -4.769           &  0.949 & 0.0552 \\
\enddata
\tablecomments{
Column Inv lists the inversion method, $\tau_{\max}$ is in years, and $p_{\max}$ accounts for the scanned lag window $\tau\in[-99,99]$ CR.
$\dagger$\ marks $|\tau_{\max}|>90$ CR.
}
\end{deluxetable}

\begin{deluxetable}{lrrrrrr}
\tablecaption{Cross-correlation summary for the cumulative longitudinal displacement $\Delta s$ at a radius of $0.997\,R_\odot$ \label{tab:lagcorr_0997}}
\tablehead{
\colhead{Inv} &
\colhead{lat} &
\colhead{$r_0$} &
\colhead{$p_{r0}$} &
\colhead{$\tau_{\max}$} &
\colhead{$r(\tau_{\max})$} &
\colhead{$p_{\max}$}}
\startdata
OLA &   0.0 & -0.165 & 0.7061 & -7.615$\dagger$   &  0.845 & 0.0830 \\
RLS &   0.0 & -0.144 & 0.7522 & -7.615$\dagger$   &  0.829 & 0.1176 \\
OLA &  15.0 &  0.634 & 0.4423 & -6.462            & -0.975 & 0.0002 \\
RLS &  15.0 &  0.659 & 0.4031 & -6.385            & -0.974 & 0.0002 \\
OLA & -15.0 &  0.361 & 0.5663 & -4.462            & -0.969 & 0.0010 \\
RLS & -15.0 &  0.363 & 0.5407 & -4.385            & -0.962 & 0.0028 \\
OLA &  30.0 &  0.680 & 0.2192 & -1.385            &  0.899 & 0.4957 \\
RLS &  30.0 &  0.691 & 0.2110 & -1.385            &  0.897 & 0.5223 \\
OLA & -30.0 &  0.624 & 0.2891 & -1.538            &  0.962 & 0.0540 \\
RLS & -30.0 &  0.655 & 0.2561 & -1.538            &  0.965 & 0.0422 \\
OLA &  45.0 & -0.779 & 0.1248 &  7.615$\dagger$   &  0.899 & 0.3471 \\
RLS &  45.0 & -0.597 & 0.2565 &  7.385$\dagger$   &  0.799 & 0.5391 \\
OLA & -45.0 & -0.728 & 0.0300 & -7.077$\dagger$   &  0.925 & 0.0154 \\
RLS & -45.0 & -0.617 & 0.1964 & -7.308$\dagger$   &  0.924 & 0.0138 \\
OLA &  60.0 & -0.126 & 0.9038 &  3.615            & -0.971 & 0.0036 \\
RLS &  60.0 & -0.158 & 0.8746 &  3.538            & -0.970 & 0.0040 \\
OLA & -60.0 &  0.401 & 0.2995 &  5.769            & -0.795 & 0.1614 \\
RLS & -60.0 &  0.477 & 0.2120 &  5.692            & -0.794 & 0.1702 \\
OLA &  67.5 &  0.479 & 0.6537 &  4.538            & -0.982 & 0.0498 \\
RLS &  67.5 &  0.526 & 0.6145 &  4.615            & -0.983 & 0.0314 \\
OLA & -67.5 &  0.007 & 0.9938 & -7.615$\dagger$   & -0.929 & 0.0696 \\
RLS & -67.5 &  0.128 & 0.8084 & -7.615$\dagger$   & -0.946 & 0.0210 \\
OLA &  75.0 &  0.741 & 0.3495 & -7.462$\dagger$   & -0.979 & 0.0008 \\
RLS &  75.0 &  0.800 & 0.2603 & -7.154$\dagger$   & -0.939 & 0.2661 \\
OLA & -75.0 & -0.596 & 0.4175 & -5.385            &  0.958 & 0.0118 \\
RLS & -75.0 & -0.648 & 0.3313 & -5.385            &  0.918 & 0.0806 \\
\enddata
\tablecomments{
Column Inv lists the inversion method, $\tau_{\max}$ is in years, and $p_{\max}$ accounts for the scanned lag window $\tau\in[-99,99]$ CR. $\dagger$\ marks $|\tau_{\max}|>90$ CR.
}
\end{deluxetable}

\begin{deluxetable}{lrrrrrr}
\tablewidth{1.0\columnwidth}
\tablecaption{Cross-correlation summary for the cumulative longitudinal displacement $\Delta s$ at radius $0.9985\,R_\odot$\label{tab:lagcorr_09985}}
\tablehead{
\colhead{Inv} &
\colhead{lat} &
\colhead{$r_0$} &
\colhead{$p_{r0}$} &
\colhead{$\tau_{\max}$} &
\colhead{$r(\tau_{\max})$} &
\colhead{$p_{\max}$}
}
\tabletypesize{\scriptsize}
\startdata
OLA &   0.0 & -0.191 & 0.6765 & -7.615$\dagger$ &  0.866 & 0.0438 \\
RLS &   0.0 & -0.207 & 0.6593 & -7.615$\dagger$ &  0.852 & 0.0686 \\
OLA & -15.0 &  0.387 & 0.5141 & -4.462 & -0.966 & 0.0014 \\
RLS & -15.0 &  0.455 & 0.3549 & -4.615 & -0.957 & 0.0062 \\
OLA &  15.0 &  0.578 & 0.5067 & -6.385 & -0.967 & 0.0012 \\
RLS &  15.0 &  0.673 & 0.3277 & -6.462 & -0.974 & 0.0002 \\
OLA & -30.0 &  0.596 & 0.3059 & -1.615 &  0.954 & 0.0800 \\
RLS & -30.0 &  0.630 & 0.2673 & -1.538 &  0.963 & 0.0476 \\
OLA &  30.0 &  0.688 & 0.2464 & -1.538 &  0.930 & 0.3139 \\
RLS &  30.0 &  0.716 & 0.2010 & -1.385 &  0.919 & 0.4051 \\
OLA & -45.0 & -0.701 & 0.0440 & -7.077$\dagger$ &  0.922 & 0.0144 \\
RLS & -45.0 & -0.559 & 0.3149 & -7.385$\dagger$ &  0.937 & 0.0062 \\
OLA &  45.0 & -0.757 & 0.2040 &  7.615$\dagger$ &  0.900 & 0.3437 \\
RLS &  45.0 & -0.755 & 0.2022 &  7.615$\dagger$ &  0.900 & 0.3363 \\
OLA & -60.0 &  0.327 & 0.4199 &  5.615 & -0.792 & 0.1644 \\
RLS & -60.0 &  0.487 & 0.2066 &  5.538 & -0.819 & 0.0918 \\
OLA &  60.0 & -0.117 & 0.9118 &  3.615 & -0.971 & 0.0030 \\
RLS &  60.0 & -0.136 & 0.8948 &  3.538 & -0.969 & 0.0060 \\
OLA & -67.5 & -0.150 & 0.8622 &  3.462 & -0.855 & 0.5351 \\
RLS & -67.5 & -0.005 & 0.9932 & -7.615$\dagger$ & -0.940 & 0.0286 \\
OLA &  67.5 &  0.402 & 0.7107 &  4.538 & -0.981 & 0.0768 \\
RLS &  67.5 &  0.443 & 0.6777 &  4.538 & -0.979 & 0.0780 \\
OLA & -75.0 & -0.624 & 0.3629 & -5.385 &  0.893 & 0.2517 \\
RLS & -75.0 & -0.598 & 0.4251 & -5.462 &  0.966 & 0.0040 \\
OLA &  75.0 &  0.741 & 0.3579 & -7.385$\dagger$ & -0.982 & 0.0006 \\
RLS &  75.0 &  0.782 & 0.2919 & -7.462$\dagger$ & -0.979 & 0.0006 \\
\enddata
\tablecomments{
Column Inv lists the inversion method, $\tau_{\max}$ is in years, and $p_{\max}$ accounts for the scanned lag window $\tau\in[-99,99]$ CR. $\dagger$\ marks $|\tau_{\max}|>90$ CR.
}
\end{deluxetable}

\bibliography{main}{}

@misc{Sun_and_Bobra,
  author = {Sun, Xudong and Bobra, Monica},
  title  = {HMI Polar Field},
  year   = {2026},
  month  = may,
  url    = {http://jsoc.stanford.edu/data/hmi/polarfield/},
  note   = {Accessed 2026 May 9}
}

@ARTICLE{Sun2015,
       author = {{Sun}, Xudong and {Hoeksema}, J. Todd and {Liu}, Yang and {Zhao}, Junwei},
        title = "{On Polar Magnetic Field Reversal and Surface Flux Transport During Solar Cycle 24}",
      journal = {\apj},
         year = 2015,
        month = jan,
       volume = {798},
       number = {2},
          eid = {114},
        pages = {114},
          doi = {10.1088/0004-637X/798/2/114},
archivePrefix = {arXiv},
       eprint = {1410.8867},
 primaryClass = {astro-ph.SR},
       adsurl = {https://ui.adsabs.harvard.edu/abs/2015ApJ...798..114S}
}

@ARTICLE{rs1999,
       author = {{Rabello-Soares}, M.~C. and {Basu}, Sarbani and {Christensen-Dalsgaard}, J.},
        title = "{On the choice of parameters in solar-structure inversion}",
      journal = {\mnras},
         year = 1999,
        month = oct,
       volume = {309},
       number = {1},
        pages = {35-47},
          doi = {10.1046/j.1365-8711.1999.02785.x},
archivePrefix = {arXiv},
       eprint = {astro-ph/9905107},
 primaryClass = {astro-ph},
       adsurl = {https://ui.adsabs.harvard.edu/abs/1999MNRAS.309...35R}
}

@ARTICLE{bogart2023,
       author = {{Bogart}, Richard S. and {Baldner}, Charles S. and {Basu}, Sarbani and {Howe}, Rachel and {Rabello Soares}, Maria Cristina},
        title = "{Evidence of a Quasiperiodic Global-scale Oscillation in the Near-surface Shear Layer of the Sun}",
      journal = {\apjl},
         year = 2023,
        month = jun,
       volume = {950},
       number = {2},
          eid = {L21},
        pages = {L21},
          doi = {10.3847/2041-8213/acd93f},
archivePrefix = {arXiv},
       eprint = {2305.18613},
 primaryClass = {astro-ph.SR},
       adsurl = {https://ui.adsabs.harvard.edu/abs/2023ApJ...950L..21B}
}

@ARTICLE{rabello2024,
       author = {{Rabello Soares}, M. Cristina and {Basu}, Sarbani and {Bogart}, Richard S.},
        title = "{Exploring the Substructure of the Near-surface Shear Layer of the Sun}",
      journal = {\apj},
         year = 2024,
        month = jun,
       volume = {967},
       number = {2},
          eid = {143},
        pages = {143},
          doi = {10.3847/1538-4357/ad3d59},
archivePrefix = {arXiv},
       eprint = {2404.02321},
 primaryClass = {astro-ph.SR},
       adsurl = {https://ui.adsabs.harvard.edu/abs/2024ApJ...967..143R}
}

@ARTICLE{hill1988,
       author = {{Hill}, Frank},
        title = "{Rings and Trumpets---Three-dimensional Power Spectra of Solar Oscillations}",
      journal = {\apj},
         year = 1988,
        month = oct,
       volume = {333},
        pages = {996},
          doi = {10.1086/166807},
       adsurl = {https://ui.adsabs.harvard.edu/abs/1988ApJ...333..996H}
}

@ARTICLE{antia2022,
       author = {{Antia}, H.~M. and {Basu}, Sarbani},
        title = "{Changes in the Near-surface Shear Layer of the Sun}",
      journal = {\apj},
         year = 2022,
        month = jan,
       volume = {924},
       number = {1},
          eid = {19},
        pages = {19},
          doi = {10.3847/1538-4357/ac32c3},
archivePrefix = {arXiv},
       eprint = {2110.13952},
 primaryClass = {astro-ph.SR},
       adsurl = {https://ui.adsabs.harvard.edu/abs/2022ApJ...924...19A}
}

@INPROCEEDINGS{Bogart_pipeline,
       author = {{Bogart}, R.~S. and {Baldner}, C. and {Basu}, S. and {Haber}, D.~A. and {Rabello-Soares}, M.~C.},
        title = "{HMI ring diagram analysis I. The processing pipeline}",
    booktitle = {GONG-SoHO 24: A New Era of Seismology of the Sun and Solar-Like Stars},
         year = 2011,
       series = {Journal of Physics Conference Series},
       volume = {271},
        month = jan,
          eid = {012008},
        pages = {012008},
          doi = {10.1088/1742-6596/271/1/012008},
       adsurl = {https://ui.adsabs.harvard.edu/abs/2011JPhCS.271a2008B}
}

@ARTICLE{basu_etal1999,
       author = {{Basu}, Sarbani and {Antia}, H.~M. and {Tripathy}, S.~C.},
        title = "{Ring Diagram Analysis of Near-Surface Flows in the Sun}",
      journal = {\apj},
         year = 1999,
        month = feb,
       volume = {512},
       number = {1},
        pages = {458-470},
          doi = {10.1086/306765},
archivePrefix = {arXiv},
       eprint = {astro-ph/9809309},
 primaryClass = {astro-ph},
       adsurl = {https://ui.adsabs.harvard.edu/abs/1999ApJ...512..458B}
}

@ARTICLE{barekat2014,
       author = {{Barekat}, A. and {Schou}, J. and {Gizon}, L.},
        title = "{The radial gradient of the near-surface shear layer of the Sun}",
      journal = {\aap},
         year = 2014,
        month = oct,
       volume = {570},
          eid = {L12},
        pages = {L12},
          doi = {10.1051/0004-6361/201424839},
archivePrefix = {arXiv},
       eprint = {1410.3162},
 primaryClass = {astro-ph.SR},
       adsurl = {https://ui.adsabs.harvard.edu/abs/2014A&A...570L..12B}
}

@ARTICLE{barekat2016,
       author = {{Barekat}, A. and {Schou}, J. and {Gizon}, L.},
        title = "{Solar-cycle variation of the rotational shear near the solar surface}",
      journal = {\aap},
         year = 2016,
        month = oct,
       volume = {595},
          eid = {A8},
        pages = {A8},
          doi = {10.1051/0004-6361/201628673},
archivePrefix = {arXiv},
       eprint = {1608.07101},
 primaryClass = {astro-ph.SR},
       adsurl = {https://ui.adsabs.harvard.edu/abs/2016A&A...595A...8B}
}

@ARTICLE{komm2022,
       author = {{Komm}, Rudolf},
        title = "{Radial Gradient of the Solar Rotation Rate in the Near-Surface Shear Layer of the Sun}",
      journal = {Frontiers in Astronomy and Space Sciences},
         year = 2022,
        month = dec,
       volume = {9},
          eid = {428},
        pages = {428},
          doi = {10.3389/fspas.2022.1017414},
       adsurl = {https://ui.adsabs.harvard.edu/abs/2022FrASS...917414K}
}

@ARTICLE{foukal1972,
       author = {{Foukal}, Peter},
        title = "{Magnetic Coupling of the Active Chromosphere to the Solar Interior}",
      journal = {\apj},
         year = 1972,
        month = apr,
       volume = {173},
        pages = {439},
          doi = {10.1086/151435},
       adsurl = {https://ui.adsabs.harvard.edu/abs/1972ApJ...173..439F}
}

@ARTICLE{karak2016,
       author = {{Karak}, Bidya Binay and {Cameron}, Robert},
        title = "{Babcock-Leighton Solar Dynamo: The Role of Downward Pumping and the Equatorward Propagation of Activity}",
      journal = {\apj},
         year = 2016,
        month = nov,
       volume = {832},
       number = {1},
          eid = {94},
        pages = {94},
          doi = {10.3847/0004-637X/832/1/94},
archivePrefix = {arXiv},
       eprint = {1605.06224},
 primaryClass = {astro-ph.SR},
       adsurl = {https://ui.adsabs.harvard.edu/abs/2016ApJ...832...94K}
}

@ARTICLE{zhao2012,
       author = {{Zhao}, Junwei and {Nagashima}, Kaori and {Bogart}, R.~S. and {Kosovichev}, A.~G. and {Duvall}, T.~L., Jr.},
        title = "{Systematic Center-to-limb Variation in Measured Helioseismic Travel Times and its Effect on Inferences of Solar Interior Meridional Flows}",
      journal = {\apjl},
         year = 2012,
        month = apr,
       volume = {749},
       number = {1},
          eid = {L5},
        pages = {L5},
          doi = {10.1088/2041-8205/749/1/L5},
archivePrefix = {arXiv},
       eprint = {1203.1904},
 primaryClass = {astro-ph.SR},
       adsurl = {https://ui.adsabs.harvard.edu/abs/2012ApJ...749L...5Z}
}

@ARTICLE{baldner2012,
       author = {{Baldner}, Charles S. and {Schou}, Jesper},
        title = "{Effects of Asymmetric Flows in Solar Convection on Oscillation Modes}",
      journal = {\apjl},
         year = 2012,
        month = nov,
       volume = {760},
       number = {1},
          eid = {L1},
        pages = {L1},
          doi = {10.1088/2041-8205/760/1/L1},
archivePrefix = {arXiv},
       eprint = {1210.1583},
 primaryClass = {astro-ph.SR},
       adsurl = {https://ui.adsabs.harvard.edu/abs/2012ApJ...760L...1B}
}

@INCOLLECTION{rhodes1990,
       author = {{Rhodes}, Edward J. and {Cacciani}, Alessandro and {Korzennik}, Sylvain G.},
        title = "{Evidence for Radial Variations in the Equatorial Profile of the Solar Internal Angular Velocity}",
    booktitle = {Progress of Seismology of the Sun and Stars},
         year = 1990,
       editor = {{Osaki}, Yoji and {Shibahashi}, Hiromoto},
       volume = {367},
        pages = {163},
    publisher = {Springer-Verlag},
          doi = {10.1007/3-540-53091-6_77},
       adsurl = {https://ui.adsabs.harvard.edu/abs/1990LNP...367..163R}
}

@INPROCEEDINGS{sekii1997,
       author = {{Sekii}, T.},
        title = "{Internal Solar rotation}",
    booktitle = {Sounding Solar and Stellar Interiors},
         year = 1997,
       editor = {{Provost}, Janine and {Schmider}, Francois-Xavier},
       volume = {181},
        month = jan,
        pages = {ISBN0792348389},
       adsurl = {https://ui.adsabs.harvard.edu/abs/1997IAUS..181..189S}
}

@ARTICLE{flow_model,
       author = {{Basu}, Sarbani and {Antia}, H.~M.},
        title = "{Large-Scale Flows in the Solar Interior: Effect of Asymmetry in Peak Profiles}",
      journal = {\apj},
         year = 1999,
        month = nov,
       volume = {525},
       number = {1},
        pages = {517-523},
          doi = {10.1086/307900},
archivePrefix = {arXiv},
       eprint = {astro-ph/9906252},
 primaryClass = {astro-ph},
       adsurl = {https://ui.adsabs.harvard.edu/abs/1999ApJ...525..517B}
}

@ARTICLE{snodgrass,
       author = {{Snodgrass}, H.~B.},
        title = "{Separation of large-scale photospheric Doppler patterns}",
      journal = {\solphys},
         year = 1984,
        month = aug,
       volume = {94},
       number = {1},
        pages = {13-31},
          doi = {10.1007/BF00154804},
       adsurl = {https://ui.adsabs.harvard.edu/abs/1984SoPh...94...13S}
}

@ARTICLE{Beck2000,
       author = {{Beck}, John G.},
        title = "{A comparison of differential rotation measurements - (Invited Review)}",
      journal = {\solphys},
         year = 2000,
        month = jan,
       volume = {191},
       number = {1},
        pages = {47-70},
          doi = {10.1023/A:1005226402796},
       adsurl = {https://ui.adsabs.harvard.edu/abs/2000SoPh..191...47B}
}

@ARTICLE{hmi,
       author = {{Scherrer}, P.~H. and {Schou}, J. and {Bush}, R.~I. and {Kosovichev}, A.~G. and {Bogart}, R.~S. and {Hoeksema}, J.~T. and {Liu}, Y. and {Duvall}, T.~L. and {Zhao}, J. and {Title}, A.~M. and {Schrijver}, C.~J. and {Tarbell}, T.~D. and {Tomczyk}, S.},
        title = "{The Helioseismic and Magnetic Imager (HMI) Investigation for the Solar Dynamics Observatory (SDO)}",
      journal = {\solphys},
         year = 2012,
        month = jan,
       volume = {275},
       number = {1-2},
        pages = {207-227},
          doi = {10.1007/s11207-011-9834-2},
       adsurl = {https://ui.adsabs.harvard.edu/abs/2012SoPh..275..207S}
}

@ARTICLE{BasuAntia2019,
       author = {{Basu}, Sarbani and {Antia}, H.~M.},
        title = "{Changes in Solar Rotation over Two Solar Cycles}",
      journal = {\apj},
         year = 2019,
        month = sep,
       volume = {883},
       number = {1},
          eid = {93},
        pages = {93},
          doi = {10.3847/1538-4357/ab3b57},
archivePrefix = {arXiv},
       eprint = {1908.05282},
 primaryClass = {astro-ph.SR},
       adsurl = {https://ui.adsabs.harvard.edu/abs/2019ApJ...883...93B}
}

@ARTICLE{AntiaBasu2001,
       author = {{Antia}, H.~M. and {Basu}, Sarbani},
        title = "{Temporal Variations of the Solar Rotation Rate at High Latitudes}",
      journal = {\apjl},
         year = 2001,
        month = sep,
       volume = {559},
       number = {1},
        pages = {L67-L70},
          doi = {10.1086/323701},
archivePrefix = {arXiv},
       eprint = {astro-ph/0108226},
 primaryClass = {astro-ph},
       adsurl = {https://ui.adsabs.harvard.edu/abs/2001ApJ...559L..67A}
}

@ARTICLE{Howeetal2018,
       author = {{Howe}, R. and {Hill}, F. and {Komm}, R. and {Chaplin}, W.~J. and {Elsworth}, Y. and {Davies}, G.~R. and {Schou}, J. and {Thompson}, M.~J.},
        title = "{Signatures of Solar Cycle 25 in Subsurface Zonal Flows}",
      journal = {\apjl},
         year = 2018,
        month = jul,
       volume = {862},
       number = {1},
          eid = {L5},
        pages = {L5},
          doi = {10.3847/2041-8213/aad1ed},
archivePrefix = {arXiv},
       eprint = {1807.02398},
 primaryClass = {astro-ph.SR},
       adsurl = {https://ui.adsabs.harvard.edu/abs/2018ApJ...862L...5H}
}

@ARTICLE{mri,
       author = {{Vasil}, Geoffrey M. and {Lecoanet}, Daniel and {Augustson}, Kyle and {Burns}, Keaton J. and {Oishi}, Jeffrey S. and {Brown}, Benjamin P. and {Brummell}, Nicholas and {Julien}, Keith},
        title = "{The solar dynamo begins near the surface}",
      journal = {\nat},
         year = 2024,
        month = may,
       volume = {629},
       number = {8013},
        pages = {769-772},
          doi = {10.1038/s41586-024-07315-1},
archivePrefix = {arXiv},
       eprint = {2404.07740},
 primaryClass = {astro-ph.SR},
       adsurl = {https://ui.adsabs.harvard.edu/abs/2024Natur.629..769V}
}

@ARTICLE{dikpati2002,
       author = {{Dikpati}, Mausumi and {Corbard}, Thierry and {Thompson}, Michael J. and {Gilman}, Peter A.},
        title = "{Flux Transport Solar Dynamos with Near-Surface Radial Shear}",
      journal = {\apjl},
         year = 2002,
        month = aug,
       volume = {575},
       number = {1},
        pages = {L41-L45},
          doi = {10.1086/342555},
       adsurl = {https://ui.adsabs.harvard.edu/abs/2002ApJ...575L..41D}
}

@ARTICLE{mason2002,
       author = {{Mason}, J. and {Hughes}, D.~W. and {Tobias}, S.~M.},
        title = "{The Competition in the Solar Dynamo between Surface and Deep-seated {\ensuremath{\alpha}}-Effects}",
      journal = {\apjl},
         year = 2002,
        month = nov,
       volume = {580},
       number = {1},
        pages = {L89-L92},
          doi = {10.1086/345419},
       adsurl = {https://ui.adsabs.harvard.edu/abs/2002ApJ...580L..89M}
}

@ARTICLE{kapyla2006,
       author = {{K{\"a}pyl{\"a}}, P.~J. and {Korpi}, M.~J. and {Tuominen}, I.},
        title = "{Solar dynamo models with {\ensuremath{\alpha}}-effect and turbulent pumping from local 3D convection calculations}",
      journal = {Astronomische Nachrichten},
         year = 2006,
        month = nov,
       volume = {327},
       number = {9},
        pages = {884},
          doi = {10.1002/asna.200610636},
archivePrefix = {arXiv},
       eprint = {astro-ph/0606089},
 primaryClass = {astro-ph},
       adsurl = {https://ui.adsabs.harvard.edu/abs/2006AN....327..884K}
}

@ARTICLE{paradkar2019,
       author = {{Paradkar}, B.~S. and {Chitre}, S.~M. and {Krishan}, V.},
        title = "{Mean field solar surface dynamo in the presence of partially ionized plasmas and sub-surface shear layer}",
      journal = {\mnras},
         year = 2019,
        month = sep,
       volume = {488},
       number = {3},
        pages = {4329-4337},
          doi = {10.1093/mnras/stz2008},
       adsurl = {https://ui.adsabs.harvard.edu/abs/2019MNRAS.488.4329P}
}

@ARTICLE{jha2021,
       author = {{Jha}, Bibhuti Kumar and {Choudhuri}, Arnab Rai},
        title = "{A theoretical model of the near-surface shear layer of the Sun}",
      journal = {\mnras},
         year = 2021,
        month = sep,
       volume = {506},
       number = {2},
        pages = {2189-2198},
          doi = {10.1093/mnras/stab1717},
archivePrefix = {arXiv},
       eprint = {2105.14266},
 primaryClass = {astro-ph.SR},
       adsurl = {https://ui.adsabs.harvard.edu/abs/2021MNRAS.506.2189J}
}

@ARTICLE{brandenburg2005,
       author = {{Brandenburg}, Axel},
        title = "{The Case for a Distributed Solar Dynamo Shaped by Near-Surface Shear}",
      journal = {\apj},
         year = 2005,
        month = may,
       volume = {625},
       number = {1},
        pages = {539-547},
          doi = {10.1086/429584},
archivePrefix = {arXiv},
       eprint = {astro-ph/0502275},
 primaryClass = {astro-ph},
       adsurl = {https://ui.adsabs.harvard.edu/abs/2005ApJ...625..539B}
}

@ARTICLE{pipin2011,
       author = {{Pipin}, V.~V. and {Kosovichev}, A.~G.},
        title = "{The Subsurface-shear-shaped Solar {\ensuremath{\alpha}}{\ensuremath{\Omega}} Dynamo}",
      journal = {\apjl},
         year = 2011,
        month = feb,
       volume = {727},
       number = {2},
          eid = {L45},
        pages = {L45},
          doi = {10.1088/2041-8205/727/2/L45},
archivePrefix = {arXiv},
       eprint = {1011.4276},
 primaryClass = {astro-ph.SR},
       adsurl = {https://ui.adsabs.harvard.edu/abs/2011ApJ...727L..45P}
}

@ARTICLE{Komm2025,
       author = {{Komm}, Rudolf and {Howe}, Rachel},
        title = "{Solar-Cycle Variation of Large-Scale Flows in the Near-Surface Shear Layer from SC 23 to SC 26}",
      journal = {\solphys},
         year = 2025,
        month = nov,
       volume = {300},
       number = {11},
          eid = {149},
        pages = {149},
          doi = {10.1007/s11207-025-02566-1},
       adsurl = {https://ui.adsabs.harvard.edu/abs/2025SoPh..300..149K}
}

@ARTICLE{Zhao2014,
       author = {{Zhao}, Junwei and {Kosovichev}, A.~G. and {Bogart}, R.~S.},
        title = "{Solar Meridional Flow in the Shallow Interior during the Rising Phase of Cycle 24}",
      journal = {\apjl},
         year = 2014,
        month = jul,
       volume = {789},
       number = {1},
          eid = {L7},
        pages = {L7},
          doi = {10.1088/2041-8205/789/1/L7},
archivePrefix = {arXiv},
       eprint = {1406.2735},
 primaryClass = {astro-ph.SR},
       adsurl = {https://ui.adsabs.harvard.edu/abs/2014ApJ...789L...7Z}
}

@ARTICLE{gong,
       author = {{Hill}, F. and {Stark}, P.~B. and {Stebbins}, R.~T. and {Anderson}, E.~R. and {Antia}, H.~M. and {Brown}, T.~M. and {Duvall}, Jr., T.~L. and {Haber}, D.~A. and {Harvey}, J.~W. and {Hathaway}, D.~H. and {Howe}, R. and {Hubbard}, R.~P. and {Jones}, H.~P. and {Kennedy}, J.~R. and {Korzennik}, S.~G. and {Kosovichev}, A.~G. and {Leibacher}, J.~W. and {Libbrecht}, K.~G. and {Pintar}, J.~A. and {Rhodes}, Jr., E.~J. and {Schou}, J. and {Thompson}, M.~J. and {Tomczyk}, S. and {Toner}, C.~G. and {Toussaint}, R. and {Williams}, W.~E.},
        title = "{The Solar Acoustic Spectrum and Eigenmode Parameters}",
      journal = {Science},
         year = 1996,
        month = may,
       volume = {272},
       number = {5266},
        pages = {1292-1295},
          doi = {10.1126/science.272.5266.1292},
       adsurl = {https://ui.adsabs.harvard.edu/abs/1996Sci...272.1292H}
}

@ARTICLE{mdi,
       author = {{Scherrer}, P.~H. and {Bogart}, R.~S. and {Bush}, R.~I. and {Hoeksema}, J.~T. and {Kosovichev}, A.~G. and {Schou}, J. and {Rosenberg}, W. and {Springer}, L. and {Tarbell}, T.~D. and {Title}, A. and {Wolfson}, C.~J. and {Zayer}, I. and {MDI Engineering Team}},
        title = "{The Solar Oscillations Investigation - Michelson Doppler Imager}",
      journal = {\solphys},
         year = 1995,
        month = dec,
       volume = {162},
       number = {1-2},
        pages = {129-188},
          doi = {10.1007/BF00733429},
       adsurl = {https://ui.adsabs.harvard.edu/abs/1995SoPh..162..129S}
}

@misc{SILSO,
author = {{Clette}, F. and {Lefèvre}, L.},
title = {SILSO Sunspot Number V2.0},
howpublished = {https://doi.org/10.24414/qnza-ac80},
month = {07},
year = {2015},
note = {Published by WDC SILSO - Royal Observatory of Belgium (ROB)}
}
\bibliographystyle{aasjournalv7}

\end{document}